\documentclass{article}
\usepackage{arxiv}
\usepackage[utf8]{inputenc}
\usepackage[T1]{fontenc}
\usepackage{hyperref}
\usepackage{url}
\usepackage{booktabs}
\usepackage{amsfonts}
\usepackage{amsmath}
\usepackage{amssymb}
\usepackage{nicefrac}
\usepackage{microtype}
\usepackage{cleveref}
\usepackage{graphicx}
\usepackage{subcaption}
\usepackage{natbib}
\usepackage{doi}
\usepackage{bm}

\graphicspath{{figures/}}

\title{A Coupled Aeroelastic-Flight Dynamic Framework for Free-Flying Flexible Aircraft with Gust Interactions}

\author{
  Nikolaos D.~Tantaroudas\thanks{Corresponding author. Senior Researcher, ICCS.} \\
  Institute of Communications and Computer Systems (ICCS)\\
  9 Iroon Politechniou Street, Zografou, Athens 15773, Greece \\
  \texttt{nikolaos.tantaroudas@iccs.gr} \\
  \And
  Ilias Karachalios \\
  National Technical University of Athens\\
  Zografou, Athens 15780, Greece \\
}

\renewcommand{\shorttitle}{Coupled Aeroelastic-Flight Dynamic Framework}

\hypersetup{
  pdftitle={A Coupled Aeroelastic-Flight Dynamic Framework for Flexible Aircraft},
  pdfsubject={physics.flu-dyn, cs.CE},
  pdfauthor={N.D. Tantaroudas, I. Karachalios},
  pdfkeywords={Aeroelasticity, Flight dynamics, Flexible aircraft, Nonlinear beam, Strip theory, Quaternions}
}

\begin{document}
\maketitle

\begin{abstract}
A complete, self-contained mathematical framework for modelling the coupled aeroelastic and flight dynamic behaviour of free-flying flexible aircraft subject to atmospheric gust encounters is presented. The framework integrates three physical disciplines: geometrically-exact nonlinear beam theory for structural dynamics, unsteady two-dimensional strip aerodynamics based on Theodorsen thin-aerofoil theory with indicial functions for shed-wake and gust-penetration effects, and quaternion-based rigid-body flight dynamics for singularity-free attitude propagation. The coupled system is assembled into a first-order state-space form amenable to time-domain simulation, model order reduction, and control design. Detailed derivations of all coupling terms, including coordinate transformations between aerodynamic and structural frames, the Jacobian block structure, and gust input matrices, are provided. Two gust models are treated: the certification-standard discrete gust and the Von K\'arm\'an continuous turbulence spectrum. The framework is verified against published benchmarks, including high-altitude long-endurance aircraft configurations and a very flexible flying-wing, demonstrating close agreement in structural frequencies, flutter speed, and static aeroelastic deflections. This paper serves as a self-contained reference for researchers implementing coupled aeroelastic-flight dynamic analysis tools for very flexible aircraft~\citep{Tantaroudas2017bookchapter, Tantaroudas2015scitech, DaRonch2014scitech_flight}.
\end{abstract}


\section{Introduction}
\label{sec:intro}

The analysis of very flexible aircraft requires an integrated treatment of structural dynamics, unsteady aerodynamics, and rigid-body flight dynamics. Traditional aerospace practice treats these disciplines separately: structural dynamics and aeroelasticity are analysed for a fixed flight condition, while flight dynamics assumes a rigid airframe~\citep{Dowell2004}. This separation is justified for conventional aircraft with small structural deformations, but breaks down for next-generation high-altitude long-endurance (HALE) and solar-powered platforms where wing deformations can exceed 25\% of the span~\citep{Patil2001, Noll2004}.

The loss of the NASA Helios prototype in 2003~\citep{Noll2004} provided dramatic evidence that coupled nonlinear analysis is essential. The mishap investigation concluded that the encounter with moderate turbulence at high dihedral angles led to a divergent pitch oscillation that the autopilot could not arrest. The structural flexibility of the wing, the resulting changes in vehicle aerodynamics and inertia distribution, and the atmospheric disturbance all played interacting roles that could not have been predicted by decoupled models.

Since the Helios accident, several research groups have developed integrated toolboxes for the analysis of very flexible aircraft. \citet{SuCesnik2010} developed UM/NAST, coupling a geometrically-exact beam with a finite-state aerodynamic model. \citet{Murua2012} presented SHARPy, which combines a displacement-based beam solver with an unsteady vortex-lattice method. \citet{Hesse2014} developed a framework at Imperial College London using an intrinsic beam formulation coupled to two-dimensional aerodynamics. At the University of Liverpool, the present authors developed the framework described in this paper, coupling a displacement-based geometrically-exact beam, unsteady strip aerodynamics, and quaternion-based flight dynamics~\citep{DaRonch2014scitech_flight, Tantaroudas2017bookchapter}.

Despite the availability of these tools, the literature lacks a single, self-contained presentation of the complete mathematical formulation with sufficient detail for independent implementation. Existing references typically present the formulation across multiple papers, use different notation conventions, or omit implementation details of the coupling terms. This paper addresses that gap by providing, in one document, the full mathematical derivation of each discipline, the coupling procedure, the Jacobian block structure, and the verification against established benchmarks.

The contributions of this paper are as follows. First, a complete formulation of the coupled structural-aerodynamic-flight dynamic system is presented in a single document, with sufficient detail for independent implementation. Second, explicit derivations of all coordinate transformations and coupling matrices that link the structural, aerodynamic, and rigid-body subsystems are provided. Third, the detailed Jacobian block structure for the assembled first-order system is documented, identifying each sub-block and its physical origin. Fourth, a comprehensive treatment of gust input modelling is given, including the 1-minus-cosine discrete gust and the Von K\'arm\'an continuous turbulence spectrum, together with the spanwise gust penetration effect. Finally, verification against published benchmarks is presented, including structural natural frequencies, flutter speed, static aeroelastic trim deflections, and mesh convergence studies that confirm the adequacy of the spatial discretisation.

The formulation has been used as the foundation for nonlinear model order reduction~\citep{DaRonch2013control, DaRonch2013gust, Tantaroudas2015scitech} and active control design~\citep{Tantaroudas2014aviation, DaRonch2014flutter, Papatheou2013ifasd} in the authors' previous work. A parametric investigation of nonlinear flexibility effects on the flight dynamics of high-aspect-ratio wings, including trim, flutter, and gust response, is reported in~\citep{Tantaroudas2026flexibility}. A model reference adaptive control framework for gust load alleviation of nonlinear aeroelastic systems is developed in~\citep{Tantaroudas2026mrac}, while foundational results on the minimum number of control laws required for nonlinear systems with input-output linearisation singularities are established in~\citep{Tantaroudas2026ballbeam}. A self-contained derivation of the NMOR formulation, including third-order Taylor expansion terms, is presented in the companion paper~\citep{Tantaroudas2026nmor}. The application of the ROM framework to rapid worst-case gust identification and $\mathcal{H}_\infty$ robust gust load alleviation are respectively demonstrated in~\citep{Tantaroudas2026gust} and~\citep{Tantaroudas2026hinf}. More recently, \citet{Artola2021} demonstrated aeroelastic control using a minimal nonlinear modal description; \citet{Goizueta2022} introduced adaptive sampling for parametric ROM interpolation; \citet{Riso2023} systematically assessed low-order modelling accuracy for very flexible wings; \citet{DelCarre2019} developed efficient time-domain simulation techniques; and a comprehensive monograph on coupled flight mechanics, aeroelasticity, and control has been provided by \citet{PalaciosCesnik2023}.

The remainder of this paper is organised as follows. Section~\ref{sec:reference_frames} defines the reference frames and coordinate systems. Section~\ref{sec:structural} presents the structural dynamics formulation, including beam kinematics, equations of motion, and finite element discretisation. Section~\ref{sec:aero} describes the unsteady strip aerodynamic model. Section~\ref{sec:flight} derives the rigid-body flight dynamic equations with quaternion kinematics. Section~\ref{sec:coupling} details the coupled system assembly, including the state vector definition, first-order state-space form, and Jacobian block structure. Section~\ref{sec:gust} presents the gust models. Section~\ref{sec:integration} describes the time integration scheme. Section~\ref{sec:verification} provides verification results. Section~\ref{sec:discussion} discusses the modelling assumptions and their implications. Conclusions are given in Section~\ref{sec:conclusions}.

\section{Reference Frames and Coordinate Systems}
\label{sec:reference_frames}

The formulation requires several coordinate frames, and the correct definition of transformations between them is essential for a consistent coupled model. The following frames are used throughout this paper (see Figure~\ref{fig:flight_frame}):

\begin{figure}[htbp]
\centering
\includegraphics[width=0.75\textwidth]{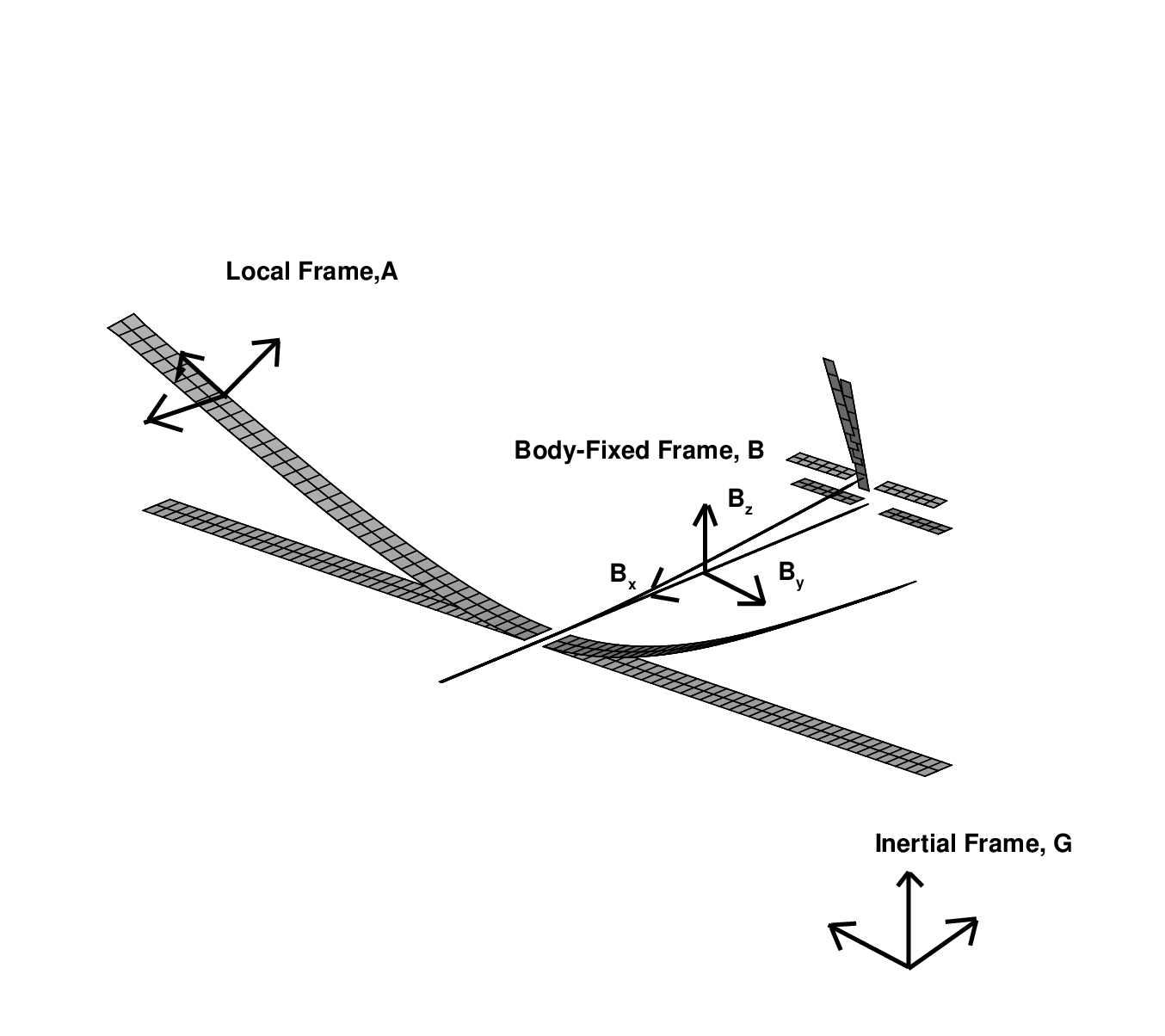}
\caption{Reference frames used in the formulation: global/inertial frame $G$, body-fixed frame $B_0$, local beam cross-section frame $B$, and local aerodynamic frame $A$.}
\label{fig:flight_frame}
\end{figure}

The global (inertial) frame $G$ is a right-handed, Earth-fixed frame with the $z_G$-axis pointing vertically downward (consistent with aerospace convention); positions and velocities for navigation are expressed in this frame. The body-fixed frame $B_0$ is attached to a reference point on the aircraft (typically the centre of mass or fuselage mid-point), with the $x_{B_0}$-axis pointing forward along the fuselage reference line, $y_{B_0}$ to starboard, and $z_{B_0}$ downward; the rigid-body flight dynamic equations are written in this frame. At each point $s$ along the beam reference line, a local beam cross-section frame $B$ is defined by the deformed cross-section orientation, with the $x_B$-axis tangent to the deformed beam centreline and $y_B$, $z_B$ defining the cross-section plane; the rotation from $G$ to $B$ is described by the rotation tensor $\mathbf{C}_{BG}(s,t)$. At each aerodynamic strip, a local aerodynamic frame $A$ is defined such that the $x_A$-axis is aligned with the local freestream velocity (projected into the cross-section plane), the $z_A$-axis is normal to the lifting surface, and $y_A$ is along the span; aerodynamic forces are first computed in this frame and then transformed to the structural frame.

The transformation between the global frame and the body-fixed frame is given by the rotation matrix $\mathbf{R}_\zeta$, which is parametrised by quaternions (Section~\ref{sec:flight}). The transformation between the body-fixed frame and the local beam frame involves both the rigid-body rotation and the elastic deformation of the beam. The transformation from the aerodynamic frame to the beam frame is the rotation $\mathbf{R}_c$ that accounts for the local angle of attack and sideslip.

\section{Structural Dynamics: Geometrically-Exact Beam Theory}
\label{sec:structural}

The aircraft structure is modelled as a system of one-dimensional beams undergoing arbitrarily large displacements and rotations~\citep{Hodges2003, Palacios2010}. This section presents the beam kinematics, strain measures, constitutive law, equations of motion, and finite element discretisation.

\subsection{Beam Kinematics}
\label{sec:kinematics}

\begin{figure}[htbp]
\centering
\includegraphics[width=0.65\textwidth]{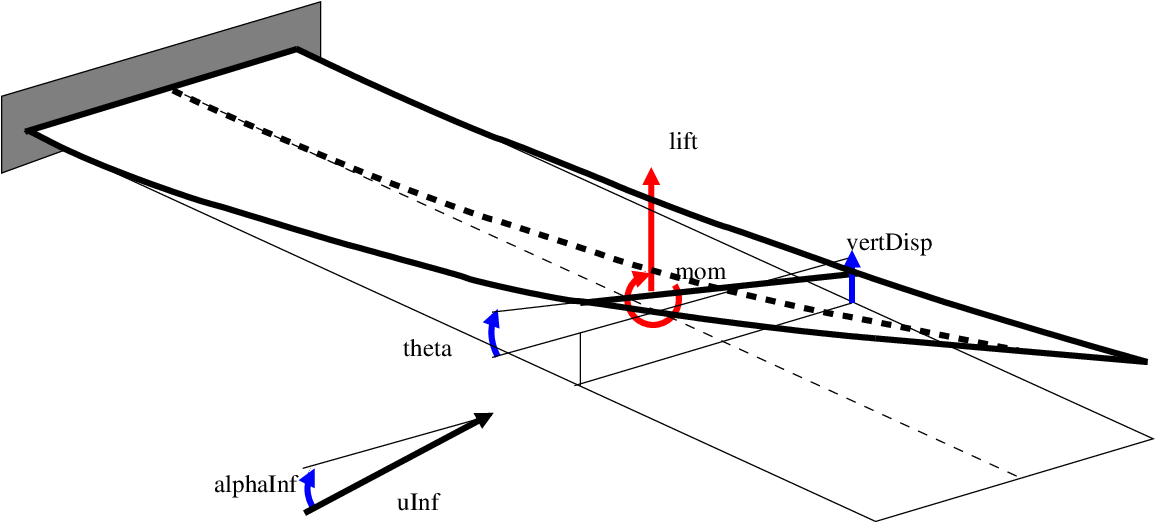}
\caption{Beam kinematics: undeformed and deformed configurations showing the displacement field $\mathbf{u}(s,t)$, the local beam frame $B$, and the strain measures $\boldsymbol{\gamma}$ and $\boldsymbol{\kappa}$.}
\label{fig:wing_deformation}
\end{figure}

Consider a straight or initially curved beam of length $L$ with curvilinear coordinate $s \in [0, L]$ measured along the undeformed reference line (see Figure~\ref{fig:wing_deformation} for an illustration of the beam kinematics). The position of a material point on the undeformed beam centreline is $\mathbf{R}_0(s)$, and the local cross-section orientation in the undeformed configuration is described by the rotation tensor $\mathbf{C}_{BG,0}(s)$.

In the deformed configuration, the position of the beam centreline becomes
\begin{equation}
\mathbf{R}(s, t) = \mathbf{R}_0(s) + \mathbf{u}(s, t),
\label{eq:beam_position}
\end{equation}
where $\mathbf{u}(s, t) = \{u_x, u_y, u_z\}^T$ is the displacement vector expressed in the global frame $G$. The local beam cross-section orientation in the deformed configuration is described by the rotation tensor $\mathbf{C}_{BG}(s, t)$, which maps vectors from the global frame $G$ to the local beam frame $B$:
\begin{equation}
\mathbf{a}_B = \mathbf{C}_{BG} \, \mathbf{a}_G,
\end{equation}
for any vector $\mathbf{a}$.

The rotation tensor $\mathbf{C}_{BG}$ belongs to the special orthogonal group $SO(3)$ and satisfies $\mathbf{C}_{BG}^T \mathbf{C}_{BG} = \mathbf{I}$ with $\det(\mathbf{C}_{BG}) = 1$. In the finite element implementation, the rotation is parametrised by Cartesian rotation vectors (also known as the Rodrigues parametrisation). Given a rotation vector $\boldsymbol{\psi} = \psi \, \hat{\mathbf{n}}$, where $\psi = \|\boldsymbol{\psi}\|$ is the rotation angle and $\hat{\mathbf{n}}$ is the unit rotation axis, the rotation tensor is computed from the Rodrigues formula:
\begin{equation}
\mathbf{C}_{BG}(\boldsymbol{\psi}) = \mathbf{I} + \frac{\sin\psi}{\psi}\widetilde{\boldsymbol{\psi}} + \frac{1 - \cos\psi}{\psi^2}\widetilde{\boldsymbol{\psi}}^2,
\label{eq:rodrigues}
\end{equation}
where $\widetilde{\boldsymbol{\psi}}$ is the skew-symmetric matrix associated with $\boldsymbol{\psi}$:
\begin{equation}
\widetilde{\boldsymbol{\psi}} = \begin{bmatrix} 0 & -\psi_3 & \psi_2 \\ \psi_3 & 0 & -\psi_1 \\ -\psi_2 & \psi_1 & 0 \end{bmatrix}.
\label{eq:skew}
\end{equation}

The angular velocity of the cross-section, expressed in the local beam frame $B$, is
\begin{equation}
\boldsymbol{\Omega}_B = \mathbf{T}(\boldsymbol{\psi}) \, \dot{\boldsymbol{\psi}},
\label{eq:angular_velocity}
\end{equation}
where the tangent operator $\mathbf{T}(\boldsymbol{\psi})$ relates the time derivative of the rotation vector to the angular velocity. For the Cartesian rotation vector parametrisation:
\begin{equation}
\mathbf{T}(\boldsymbol{\psi}) = \mathbf{I} + \frac{1 - \cos\psi}{\psi^2}\widetilde{\boldsymbol{\psi}} + \frac{\psi - \sin\psi}{\psi^3}\widetilde{\boldsymbol{\psi}}^2.
\label{eq:tangent_operator}
\end{equation}

\subsection{Strain Measures}
\label{sec:strains}

The force strain measure $\boldsymbol{\gamma}$ and the moment strain measure $\boldsymbol{\kappa}$ are defined in the local beam frame $B$ as~\citep{Hodges2003}:
\begin{equation}
\boldsymbol{\gamma}(s,t) = \mathbf{C}_{BG}(s,t) \, \mathbf{R}'(s,t) - \mathbf{e}_1,
\label{eq:gamma}
\end{equation}
\begin{equation}
\boldsymbol{\kappa}(s,t) = \mathbf{K}(s,t) - \mathbf{K}_0(s),
\label{eq:kappa}
\end{equation}
where $(\cdot)' = \partial(\cdot)/\partial s$ denotes the derivative with respect to the arc-length coordinate, $\mathbf{e}_1 = \{1, 0, 0\}^T$ is the unit vector along the undeformed beam tangent, $\mathbf{K}(s,t)$ is the curvature vector in the deformed configuration, and $\mathbf{K}_0(s)$ is the initial curvature vector.

The curvature vector $\mathbf{K}$ is the axial vector of the skew-symmetric matrix $\widetilde{\mathbf{K}} = \mathbf{C}_{BG} \, \mathbf{C}_{BG}'^T$, which can be written as
\begin{equation}
\widetilde{\mathbf{K}}(s,t) = \mathbf{C}_{BG}(s,t) \, \frac{\partial \mathbf{C}_{BG}^T(s,t)}{\partial s}.
\label{eq:curvature_tensor}
\end{equation}

The force strain $\boldsymbol{\gamma}$ has three components: $\gamma_1$ represents axial extension, while $\gamma_2$ and $\gamma_3$ represent transverse shear strains. The moment strain $\boldsymbol{\kappa}$ has components: $\kappa_1$ is the twist, while $\kappa_2$ and $\kappa_3$ are the bending curvatures about the two cross-section principal axes.

For an initially straight and untwisted beam ($\mathbf{K}_0 = \mathbf{0}$) with no pre-strain, the undeformed state corresponds to $\boldsymbol{\gamma} = \mathbf{0}$ and $\boldsymbol{\kappa} = \mathbf{0}$.

\subsection{Constitutive Law}
\label{sec:constitutive}

The cross-sectional internal forces $\mathbf{F}_s = \{F_1, F_2, F_3\}^T$ (axial force and two shear forces) and internal moments $\mathbf{M}_s = \{M_1, M_2, M_3\}^T$ (torque and two bending moments) are related to the strain measures through the cross-sectional stiffness matrix $\mathbf{S}$:
\begin{equation}
\begin{Bmatrix} \mathbf{F}_s \\ \mathbf{M}_s \end{Bmatrix} = \mathbf{S} \begin{Bmatrix} \boldsymbol{\gamma} \\ \boldsymbol{\kappa} \end{Bmatrix}, \qquad \mathbf{S} = \begin{bmatrix} \mathbf{S}_{11} & \mathbf{S}_{12} \\ \mathbf{S}_{12}^T & \mathbf{S}_{22} \end{bmatrix} \in \mathbb{R}^{6 \times 6}.
\label{eq:constitutive}
\end{equation}

For isotropic or symmetric cross-sections, the stiffness matrix $\mathbf{S}$ is diagonal or block-diagonal. For a general anisotropic composite cross-section, $\mathbf{S}$ is a fully populated $6 \times 6$ matrix that must be obtained from a cross-sectional analysis tool~\citep{Palacios2010intrinsic}. For many applications, the simplified diagonal form is sufficient:
\begin{equation}
\mathbf{S} = \mathrm{diag}\left(EA, \; GA_2, \; GA_3, \; GJ, \; EI_2, \; EI_3\right),
\label{eq:stiffness_diagonal}
\end{equation}
where $EA$ is the axial stiffness, $GA_2$ and $GA_3$ are the shear stiffnesses, $GJ$ is the torsional stiffness, and $EI_2$ and $EI_3$ are the bending stiffnesses about the two principal axes.

\subsection{Equations of Motion}
\label{sec:eom}

The equations of motion are derived from Hamilton's principle. For a beam element, the weak form of the equations of motion, after spatial discretisation and linearisation, takes the matrix form~\citep{Hesse2014, Tantaroudas2017bookchapter}:
\begin{equation}
\mathbf{M}(\mathbf{w}_s) \begin{Bmatrix} \ddot{\mathbf{w}}_s \\ \ddot{\mathbf{w}}_r \end{Bmatrix} + \mathbf{Q}_{gyr}\!\left(\mathbf{w}_s, \dot{\mathbf{w}}_s, \mathbf{w}_r\right) \begin{Bmatrix} \dot{\mathbf{w}}_s \\ \dot{\mathbf{w}}_r \end{Bmatrix} + \mathbf{Q}_{stiff}(\mathbf{w}_s) \begin{Bmatrix} \mathbf{w}_s \\ \mathbf{w}_r \end{Bmatrix} = \mathbf{R}_F.
\label{eq:eom}
\end{equation}

The terms in this equation have the following physical interpretations:

\paragraph{Mass matrix $\mathbf{M}$.} The tangent mass matrix depends on the structural state $\mathbf{w}_s$ for geometrically nonlinear problems (because the inertia distribution changes with deformation). It couples structural and rigid-body inertias through off-diagonal blocks:
\begin{equation}
\mathbf{M} = \begin{bmatrix} \mathbf{M}_{ss} & \mathbf{M}_{sr} \\ \mathbf{M}_{rs} & \mathbf{M}_{rr} \end{bmatrix},
\label{eq:mass_matrix}
\end{equation}
where $\mathbf{M}_{ss}$ is the structural mass matrix, $\mathbf{M}_{rr}$ is the rigid-body mass-inertia matrix, and $\mathbf{M}_{sr} = \mathbf{M}_{rs}^T$ are the coupling terms that arise because the structural nodal velocities include contributions from the rigid-body motion. For a lumped mass approach, the mass per unit length $\mu(s)$ and the rotational inertia per unit length $\mathbf{J}_s(s)$ are distributed to the finite element nodes.

The elemental mass matrix for a two-noded beam element with shape functions $N_1(s)$ and $N_2(s)$ is
\begin{equation}
\mathbf{M}^{(e)} = \int_0^{l_e} \begin{bmatrix} \mu N_1 N_1 \mathbf{I}_3 & \mathbf{0} & \mu N_1 N_2 \mathbf{I}_3 & \mathbf{0} \\ \mathbf{0} & \mathbf{J}_s N_1 N_1 & \mathbf{0} & \mathbf{J}_s N_1 N_2 \\ \mu N_2 N_1 \mathbf{I}_3 & \mathbf{0} & \mu N_2 N_2 \mathbf{I}_3 & \mathbf{0} \\ \mathbf{0} & \mathbf{J}_s N_2 N_1 & \mathbf{0} & \mathbf{J}_s N_2 N_2 \end{bmatrix} ds,
\label{eq:element_mass}
\end{equation}
where $l_e$ is the element length and $\mathbf{I}_3$ is the $3 \times 3$ identity matrix.

\paragraph{Gyroscopic terms $\mathbf{Q}_{gyr}$.} This term contains gyroscopic and Coriolis forces arising from the interaction between structural deformation rates and rigid-body angular velocity. For a free-flying aircraft with angular velocity $\boldsymbol{\omega}$, the velocity of any material point on the beam is
\begin{equation}
\mathbf{v}_{point} = \mathbf{v}_{cm} + \boldsymbol{\omega} \times \mathbf{r} + \dot{\mathbf{u}},
\label{eq:point_velocity}
\end{equation}
where $\mathbf{v}_{cm}$ is the velocity of the centre of mass, $\mathbf{r}$ is the position vector from the centre of mass, and $\dot{\mathbf{u}}$ is the structural velocity. The cross terms between $\boldsymbol{\omega}$ and $\dot{\mathbf{u}}$ give rise to the Coriolis forces, while the quadratic terms in $\boldsymbol{\omega}$ produce centrifugal stiffening. These terms are critical for the correct prediction of phugoid and short-period modes of flexible aircraft~\citep{Patil2006, Hesse2014}.

\paragraph{Elastic stiffness $\mathbf{Q}_{stiff}$.} The elastic restoring forces are computed from the constitutive law, Eq.~\eqref{eq:constitutive}. For geometrically nonlinear analysis, $\mathbf{Q}_{stiff}$ is a nonlinear function of the structural displacements and rotations. The tangent stiffness matrix, obtained by linearisation about the current deformed state, is
\begin{equation}
\mathbf{K}_{tan} = \frac{\partial \mathbf{Q}_{stiff}}{\partial \mathbf{w}_s} = \mathbf{K}_{mat} + \mathbf{K}_{geo},
\label{eq:tangent_stiffness}
\end{equation}
where $\mathbf{K}_{mat}$ is the material (constitutive) stiffness and $\mathbf{K}_{geo}$ is the geometric stiffness that accounts for the effect of internal forces on the equilibrium of the deformed configuration.

\paragraph{Applied forces $\mathbf{R}_F$.} The right-hand side contains all external forces and moments: aerodynamic loads, gravitational forces, propulsive forces, and control surface forces. The aerodynamic contributions are computed from the strip theory model described in Section~\ref{sec:aero}.

\subsection{Finite Element Discretisation}
\label{sec:fem}

\begin{figure}[htbp]
\centering
\includegraphics[width=0.7\textwidth]{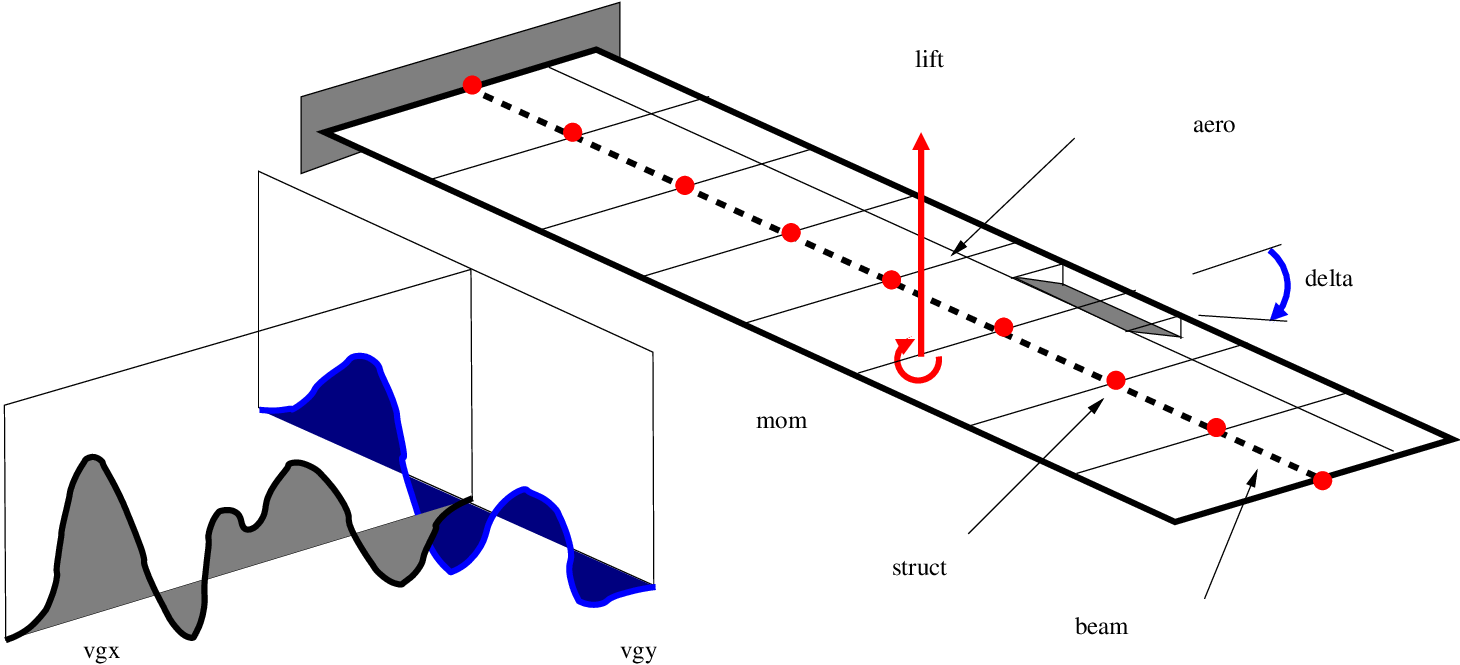}
\caption{Wing structural layout showing the finite element discretisation with two-noded beam elements, aerodynamic strip locations, and control surface segments.}
\label{fig:wing_layout}
\end{figure}

The beam is discretised using two-noded displacement-based finite elements (Figure~\ref{fig:wing_layout}). Each node carries 6 degrees of freedom: 3 translations $\{u_x, u_y, u_z\}$ and 3 rotations parametrised by the Cartesian rotation vector $\{\psi_1, \psi_2, \psi_3\}$. The element nodal displacement vector is therefore
\begin{equation}
\mathbf{q}^{(e)} = \left\{ u_{x,1}, u_{y,1}, u_{z,1}, \psi_{1,1}, \psi_{2,1}, \psi_{3,1}, \; u_{x,2}, u_{y,2}, u_{z,2}, \psi_{1,2}, \psi_{2,2}, \psi_{3,2} \right\}^T \in \mathbb{R}^{12}.
\label{eq:element_dof}
\end{equation}

The displacement and rotation fields within the element are interpolated using linear Lagrangian shape functions:
\begin{equation}
N_1(\xi) = \frac{1-\xi}{2}, \qquad N_2(\xi) = \frac{1+\xi}{2}, \qquad \xi \in [-1, 1].
\label{eq:shape_functions}
\end{equation}

The displacement field is interpolated as
\begin{equation}
\mathbf{u}(\xi) = N_1(\xi) \, \mathbf{u}_1 + N_2(\xi) \, \mathbf{u}_2,
\label{eq:displacement_interp}
\end{equation}
and the rotation vector is interpolated similarly. While linear interpolation of rotations is strictly valid only for small incremental rotations, it is adequate for the element sizes used in the present applications (typically 10--20 elements per wing semi-span), as confirmed by mesh convergence studies.

The spatial derivatives needed for the strain measures, Eqs.~\eqref{eq:gamma}--\eqref{eq:kappa}, are computed using the element Jacobian:
\begin{equation}
\frac{\partial(\cdot)}{\partial s} = \frac{2}{l_e}\frac{\partial(\cdot)}{\partial \xi}.
\label{eq:jacobian_mapping}
\end{equation}

The global system is assembled by standard finite element procedures. Appropriate boundary conditions are applied: clamped (wing root fixed to fuselage) for cantilevered wings or free-free for the complete free-flying aircraft. The assembled structural system has dimension $n_s = 6 \times N_{nodes}$ for the displacement DOFs, giving a total of $2 n_s$ first-order states when velocities are included.

Reduced integration (single Gauss point per element) is used for the evaluation of the stiffness matrices to avoid shear locking in slender beam elements. The consistent mass matrix, Eq.~\eqref{eq:element_mass}, can be replaced by a lumped mass matrix for computational efficiency without significant loss of accuracy for the low-frequency modes of interest.

\begin{figure}[htbp]
\centering
\includegraphics[width=0.75\textwidth]{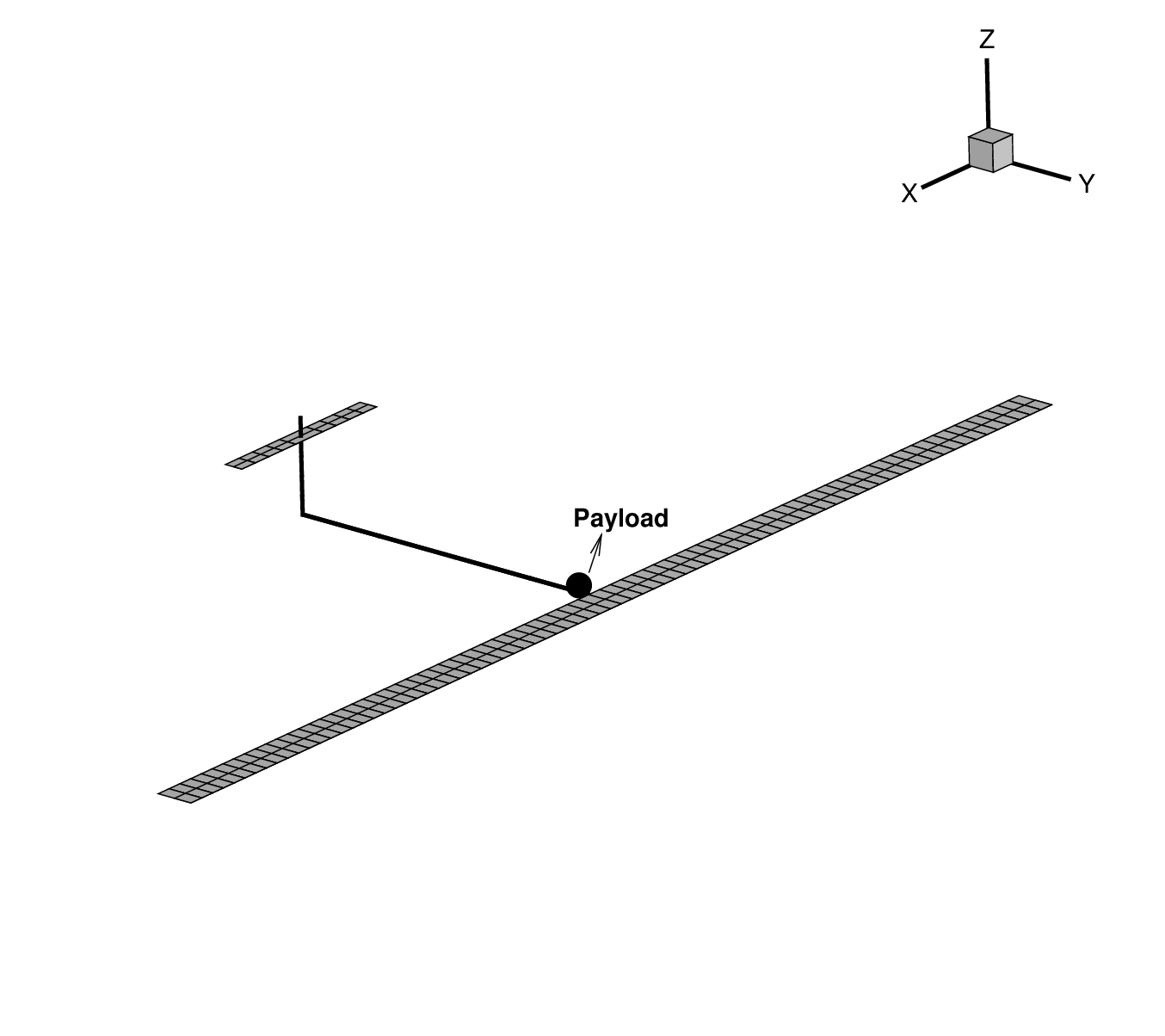}
\caption{Finite element model of the HALE aircraft showing the beam element discretisation. The wing, horizontal tail, vertical tail, and fuselage are modelled as interconnected beams with appropriate stiffness and mass properties.}
\label{fig:hale_fem}
\end{figure}

\section{Aerodynamic Model: Unsteady Strip Theory}
\label{sec:aero}

The aerodynamic loads are computed using unsteady two-dimensional strip theory based on the thin-aerofoil theory of \citet{Theodorsen1935}. At each spanwise station (strip), the local aerodynamic loads are determined from the section motion, the trailing-edge flap deflection, and the gust velocity. The loads from these three sources are computed independently and superimposed, exploiting the linearity of the thin-aerofoil theory. The time-domain formulation uses indicial (step-response) functions rather than the frequency-domain Theodorsen function, following the approach of \citet{Jones1938} and \citet{Wagner1925}.

\subsection{Section Motion Contributions: The Wagner Problem}
\label{sec:wagner}

Consider an aerofoil strip with semi-chord $b$, chord $c = 2b$, elastic axis at a distance $a \cdot b$ aft of the mid-chord (where $a = -1$ places the elastic axis at the leading edge and $a = +1$ at the trailing edge), freestream velocity $U$, plunge displacement $h$ (positive downward), and pitch angle $\alpha$ about the elastic axis.

The non-dimensional time is defined as $\tau = Ut/b$ (measured in semi-chords of travel). The effective angle of attack accounting for the quasi-steady contribution of plunge rate and pitch rate is
\begin{equation}
\alpha_{eff}(t) = \alpha(t) + \frac{\dot{h}(t)}{U} + \left(\frac{1}{2} - a\right)\frac{b\dot{\alpha}(t)}{U}.
\label{eq:alpha_eff}
\end{equation}

The total lift coefficient due to section motion is decomposed into circulatory and non-circulatory parts:
\begin{equation}
C_{L,s}(t) = C_{L,s}^{circ}(t) + C_{L,s}^{nc}(t).
\label{eq:CL_section}
\end{equation}

The circulatory lift includes the memory effect of the shed wake through the Wagner indicial function $\phi_w$:
\begin{equation}
C_{L,s}^{circ}(t) = C_{L_\alpha} \left[ \phi_w(0) \, \alpha_{eff}(t) + \int_0^t \dot{\alpha}_{eff}(\sigma) \, \phi_w(t - \sigma) \, d\sigma \right],
\label{eq:CL_circ}
\end{equation}
where $C_{L_\alpha} = 2\pi$ for a thin aerofoil. The Wagner function $\phi_w(\tau)$ represents the build-up of circulatory lift following a step change in angle of attack~\citep{Wagner1925}. It starts at $\phi_w(0) = 0.5$ and asymptotically approaches unity.

The non-circulatory (apparent mass) contributions arise from the acceleration of the fluid mass surrounding the aerofoil:
\begin{equation}
C_{L,s}^{nc}(t) = \frac{\pi b}{U}\dot{\alpha}(t) + \frac{\pi b}{U^2}\ddot{h}(t) - \frac{\pi a b^2}{U^2}\ddot{\alpha}(t).
\label{eq:CL_nc}
\end{equation}

The pitching moment coefficient about the elastic axis due to section motion is
\begin{equation}
C_{m,s}(t) = -\frac{\pi b}{2U}\dot{\alpha}(t) + \left(\frac{1}{2} + a\right) C_{L,s}^{circ}(t) \cdot \frac{b}{c} + C_{m,s}^{nc}(t),
\label{eq:Cm_section}
\end{equation}
where the non-circulatory moment includes contributions from the apparent inertia:
\begin{equation}
C_{m,s}^{nc}(t) = -\frac{\pi b^2}{U^2}\left[\frac{1}{2}\ddot{h}(t) + a b \ddot{\alpha}(t) + \frac{b}{U}\left(\frac{1}{8} - \frac{a^2}{2}\right)\dot{\alpha}(t)\right] \cdot \frac{1}{c}.
\label{eq:Cm_nc}
\end{equation}

The drag coefficient is obtained from a combination of profile drag $C_{D_0}$ and the component of lift in the drag direction due to the effective angle of attack:
\begin{equation}
C_{D,s}(t) = C_{D_0} + C_{L,s}(t) \sin\alpha_{eff}(t) \approx C_{D_0} + C_{L,s}(t) \, \alpha_{eff}(t).
\label{eq:CD_section}
\end{equation}

\subsection{Wagner Indicial Function and Augmented States}
\label{sec:wagner_function}

The Wagner indicial function is approximated by a two-term exponential series~\citep{Jones1938}:
\begin{equation}
\phi_w(\tau) = 1 - \Psi_1 e^{-\varepsilon_1 \tau} - \Psi_2 e^{-\varepsilon_2 \tau},
\label{eq:wagner_approx}
\end{equation}
with the Jones approximation constants $\Psi_1 = 0.165$, $\Psi_2 = 0.335$, $\varepsilon_1 = 0.0455$, $\varepsilon_2 = 0.3$. The value at $\tau = 0$ is $\phi_w(0) = 1 - \Psi_1 - \Psi_2 = 0.5$, consistent with the exact result.

Using the exponential approximation, the Duhamel integral in Eq.~\eqref{eq:CL_circ} can be evaluated through augmented (internal) aerodynamic states. Define two augmented states per strip:
\begin{equation}
\lambda_k^w(t) = \int_0^t \dot{\alpha}_{eff}(\sigma) \, e^{-\varepsilon_k U(t-\sigma)/b} \, d\sigma, \qquad k = 1, 2.
\label{eq:wagner_states}
\end{equation}

These satisfy the ordinary differential equations:
\begin{equation}
\dot{\lambda}_k^w(t) = \dot{\alpha}_{eff}(t) - \frac{\varepsilon_k U}{b} \lambda_k^w(t), \qquad k = 1, 2.
\label{eq:wagner_ode}
\end{equation}

The circulatory lift, Eq.~\eqref{eq:CL_circ}, is then expressed as
\begin{equation}
C_{L,s}^{circ}(t) = C_{L_\alpha} \left[ \alpha_{eff}(t) - \Psi_1 \lambda_1^w(t) - \Psi_2 \lambda_2^w(t) \right].
\label{eq:CL_circ_states}
\end{equation}

The augmented states decay exponentially in the absence of forcing, with time constants $b/(\varepsilon_k U)$, representing the memory of the shed wake.

\subsection{Trailing-Edge Flap Contributions}
\label{sec:flap}

For strips equipped with a trailing-edge control surface (flap), additional aerodynamic loads arise from flap deflection $\delta$. The flap has a chord $c_f$ and the flap-to-chord ratio is $E = c_f/c$. The hinge is located at a distance $(1-E)c$ from the leading edge.

The flap effectiveness coefficients depend on the Theodorsen constants $T_1$ through $T_{14}$~\citep{Theodorsen1935, Dowell2004}, which are geometric functions of the flap ratio $E$:
\begin{align}
T_1 &= -\frac{1}{3}\sqrt{1-E^2}(2+E^2) + E \cos^{-1}E, \\
T_2 &= E(1-E^2) - \sqrt{1-E^2}(1+E^2)\cos^{-1}E + E(\cos^{-1}E)^2, \\
T_3 &= -\left(\frac{1}{8}+E^2\right)(\cos^{-1}E)^2 + \frac{1}{4}E\sqrt{1-E^2}\cos^{-1}E \cdot (7+2E^2) \nonumber \\
&\quad - \frac{1}{8}(1-E^2)(5E^2+4).
\end{align}

The lift and pitching moment increments due to flap deflection are
\begin{align}
C_{L,f}(t) &= C_{L_\delta}\left[\phi_w(0)\,\delta(t) + \int_0^t \dot{\delta}(\sigma)\phi_w(t-\sigma)\,d\sigma\right] + C_{L,f}^{nc}(t), \label{eq:CL_flap} \\
C_{m,f}(t) &= C_{m_\delta}\left[\phi_w(0)\,\delta(t) + \int_0^t \dot{\delta}(\sigma)\phi_w(t-\sigma)\,d\sigma\right] + C_{m,f}^{nc}(t), \label{eq:Cm_flap}
\end{align}
where $C_{L_\delta}$ and $C_{m_\delta}$ are the flap lift and moment effectiveness coefficients:
\begin{equation}
C_{L_\delta} = 2\left(\cos^{-1}E + \sqrt{1-E^2}\right), \qquad C_{m_\delta} = -\frac{T_1}{2} - \frac{\pi}{4}(1-E).
\end{equation}

The flap contributions use the same Wagner indicial function as the section motion, introducing two additional augmented states per flapped strip (or sharing the same indicial function with separate input). The non-circulatory flap loads $C_{L,f}^{nc}$ and $C_{m,f}^{nc}$ depend on $\dot{\delta}$ and $\ddot{\delta}$ through the apparent-mass effect:
\begin{align}
C_{L,f}^{nc}(t) &= -\frac{b T_4}{U}\dot{\delta}(t), \\
C_{m,f}^{nc}(t) &= \frac{T_4 + T_{10}}{2}\frac{b}{U}\dot{\delta}(t) + \frac{b^2 T_{11}}{U^2}\ddot{\delta}(t),
\end{align}
where $T_4$ and $T_{10}$, $T_{11}$ are further Theodorsen constants dependent on $E$.

The hinge moment coefficient is similarly obtained:
\begin{equation}
C_{h,f}(t) = C_{h_\alpha}\alpha_{eff}(t) + C_{h_\delta}\delta(t) + C_{h,f}^{nc}(t),
\end{equation}
with coefficients expressed in terms of the Theodorsen constants. The hinge moment is required for control system sizing and actuator power estimation.

\subsection{Atmospheric Gust Contributions: The K\"ussner Problem}
\label{sec:kussner}

Gust-induced loads model the response of the aerofoil to a vertical gust field $w_g(t)$ that the strip encounters as it flies through the gust. The gust effectively changes the angle of attack by $\Delta\alpha_g = w_g(t)/U$. However, the build-up of circulatory lift due to a sharp-edged gust differs from the Wagner response to a step in angle of attack; it is governed by the K\"ussner indicial function $\Psi_k$~\citep{Kussner1936}.

The gust-induced lift coefficient is
\begin{equation}
C_{L,g}(t) = C_{L_\alpha}\frac{w_g(t)}{U}\Psi_k(0) + C_{L_\alpha}\int_0^t \frac{\dot{w}_g(\sigma)}{U}\Psi_k(t-\sigma)\,d\sigma.
\label{eq:CL_gust}
\end{equation}

The K\"ussner function is approximated by the two-term exponential~\citep{Jones1938}:
\begin{equation}
\Psi_k(\tau) = 1 - \Psi_3 e^{-\varepsilon_3 \tau} - \Psi_4 e^{-\varepsilon_4 \tau},
\label{eq:kussner_approx}
\end{equation}
with $\Psi_3 = 0.5792$, $\Psi_4 = 0.4208$, $\varepsilon_3 = 0.1393$, $\varepsilon_4 = 1.802$. The value $\Psi_k(0) = 1 - \Psi_3 - \Psi_4 = 0$ means that gust-induced circulatory lift starts from zero and builds up gradually, in contrast to the Wagner function.

Two additional augmented states per strip are introduced for the gust memory:
\begin{equation}
\lambda_k^g(t) = \int_0^t \frac{\dot{w}_g(\sigma)}{U} e^{-\varepsilon_{k+2} U(t-\sigma)/b}\,d\sigma, \qquad k = 1, 2,
\label{eq:kussner_states}
\end{equation}
satisfying the ODEs:
\begin{equation}
\dot{\lambda}_k^g(t) = \frac{\dot{w}_g(t)}{U} - \frac{\varepsilon_{k+2} U}{b}\lambda_k^g(t), \qquad k = 1, 2.
\label{eq:kussner_ode}
\end{equation}

The gust-induced lift then becomes
\begin{equation}
C_{L,g}(t) = C_{L_\alpha}\left[\frac{w_g(t)}{U} - \Psi_3 \lambda_1^g(t) - \Psi_4 \lambda_2^g(t)\right].
\label{eq:CL_gust_states}
\end{equation}

The gust-induced pitching moment about the elastic axis is
\begin{equation}
C_{m,g}(t) = \left(\frac{1}{2} + a\right)\frac{b}{c}\,C_{L,g}(t).
\label{eq:Cm_gust}
\end{equation}

\subsection{Total Aerodynamic Loads and Force Transformation}
\label{sec:aero_total}

The total aerodynamic coefficients at each strip are obtained by linear superposition:
\begin{equation}
C_i(t) = C_{i,s}(t) + C_{i,f}(t) + C_{i,g}(t), \qquad i \in \{L, D, m\}.
\label{eq:aero_superposition}
\end{equation}

The sectional aerodynamic forces and moments in the local aerodynamic frame $A$ are
\begin{equation}
\mathbf{f}_A = \frac{1}{2}\rho U_{loc}^2 c \, \Delta y \begin{Bmatrix} -C_D \\ C_L \\ 0 \end{Bmatrix}, \qquad \mathbf{m}_A = \frac{1}{2}\rho U_{loc}^2 c^2 \, \Delta y \begin{Bmatrix} C_m \\ 0 \\ 0 \end{Bmatrix},
\label{eq:aero_forces}
\end{equation}
where $\rho$ is the air density, $U_{loc}$ is the local effective freestream velocity magnitude, and $\Delta y$ is the strip width along the span.

These forces must be transformed from the aerodynamic frame $A$ to the structural beam frame $B$ for insertion into the structural equations of motion. The transformation is
\begin{equation}
\mathbf{f}_B = \mathbf{R}_c \, \mathbf{f}_A, \qquad \mathbf{m}_B = \mathbf{R}_c \, \mathbf{m}_A + \mathbf{r}_{ea} \times \mathbf{f}_B,
\label{eq:force_transform}
\end{equation}
where $\mathbf{R}_c$ is the rotation matrix from the aerodynamic frame to the beam frame (which depends on the local angle of attack and sweep) and $\mathbf{r}_{ea}$ is the offset between the aerodynamic centre (quarter-chord) and the elastic axis.

The total number of augmented aerodynamic states per strip is four: 2 states for the Wagner (section motion) memory and 2 states for the K\"ussner (gust) memory, with 2 additional states if a trailing-edge flap is present. For a model with $N_s$ strips (of which $N_f$ have flaps), the total number of aerodynamic augmented states is $n_f = 4 N_s + 2 N_f$.

\section{Rigid-Body Flight Dynamics}
\label{sec:flight}

The rigid-body flight dynamic equations describe the motion of the aircraft as a whole through space. They are written in the body-fixed frame $B_0$ to exploit the simplification that the inertia tensor (though dependent on the elastic deformation) does not change due to the rigid-body motion itself.

\subsection{Newton--Euler Equations in the Body Frame}
\label{sec:newton_euler}

The translational equations of motion in the body frame are~\citep{Patil2001, Hesse2014}:
\begin{equation}
m\left(\dot{\mathbf{v}} + \boldsymbol{\omega} \times \mathbf{v}\right) = \mathbf{F}_{aero} + \mathbf{F}_{grav} + \mathbf{F}_{thrust},
\label{eq:translation}
\end{equation}
where $m$ is the total aircraft mass, $\mathbf{v} = \{u, v, w\}^T$ is the velocity of the centre of mass expressed in the body frame, $\boldsymbol{\omega} = \{p, q, r\}^T$ is the angular velocity of the body frame, and $\mathbf{F}_{aero}$, $\mathbf{F}_{grav}$, and $\mathbf{F}_{thrust}$ are the aerodynamic, gravitational, and propulsive forces, respectively, all expressed in the body frame.

The expanded form of Eq.~\eqref{eq:translation} gives three scalar equations:
\begin{align}
m(\dot{u} + qw - rv) &= F_{aero,x} + F_{grav,x} + F_{thrust,x}, \label{eq:trans_x} \\
m(\dot{v} + ru - pw) &= F_{aero,y} + F_{grav,y} + F_{thrust,y}, \label{eq:trans_y} \\
m(\dot{w} + pv - qu) &= F_{aero,z} + F_{grav,z} + F_{thrust,z}. \label{eq:trans_z}
\end{align}

The gravitational force in the body frame is obtained by rotating the gravity vector from the global frame:
\begin{equation}
\mathbf{F}_{grav} = m \, \mathbf{R}_\zeta^T \begin{Bmatrix} 0 \\ 0 \\ g \end{Bmatrix} = m g \begin{Bmatrix} 2(\zeta_1\zeta_3 - \zeta_0\zeta_2) \\ 2(\zeta_2\zeta_3 + \zeta_0\zeta_1) \\ 1 - 2(\zeta_1^2 + \zeta_2^2) \end{Bmatrix},
\label{eq:gravity_body}
\end{equation}
where $g = 9.81$ m/s$^2$ and the quaternion components $\zeta_i$ define the attitude (Section~\ref{sec:quaternion}).

\subsection{Rotational Dynamics}
\label{sec:rotational}

The rotational equations of motion (Euler's equations) in the body frame are:
\begin{equation}
\mathbf{J}\dot{\boldsymbol{\omega}} + \boldsymbol{\omega} \times \left(\mathbf{J}\boldsymbol{\omega}\right) = \mathbf{M}_{aero} + \mathbf{M}_{thrust},
\label{eq:rotation}
\end{equation}
where $\mathbf{J}$ is the inertia tensor of the deformed aircraft about the centre of mass and $\mathbf{M}_{aero}$ and $\mathbf{M}_{thrust}$ are the aerodynamic and propulsive moments.

For a flexible aircraft, the inertia tensor $\mathbf{J}$ depends on the structural deformation:
\begin{equation}
\mathbf{J} = \sum_{i=1}^{N_{nodes}} m_i \left[\left(\mathbf{r}_i^T \mathbf{r}_i\right)\mathbf{I}_3 - \mathbf{r}_i \mathbf{r}_i^T\right] + \mathbf{J}_{s,i},
\label{eq:inertia_tensor}
\end{equation}
where $m_i$ is the lumped mass at node $i$, $\mathbf{r}_i$ is the position of node $i$ relative to the centre of mass (in the deformed configuration), and $\mathbf{J}_{s,i}$ is the rotational inertia of the cross-section at node $i$ (transformed to the body frame). The dependence of $\mathbf{J}$ on the structural state is a key coupling mechanism between the structural and flight dynamic subsystems.

In expanded scalar form, Eq.~\eqref{eq:rotation} gives:
\begin{align}
J_{xx}\dot{p} - J_{xy}\dot{q} - J_{xz}\dot{r} + (J_{zz} - J_{yy})qr - J_{xz}pq + J_{xy}pr &= M_{aero,x}, \label{eq:rot_x} \\
-J_{xy}\dot{p} + J_{yy}\dot{q} - J_{yz}\dot{r} + (J_{xx} - J_{zz})pr + J_{yz}pq - J_{xy}qr &= M_{aero,y}, \label{eq:rot_y} \\
-J_{xz}\dot{p} - J_{yz}\dot{q} + J_{zz}\dot{r} + (J_{yy} - J_{xx})pq + J_{xz}qr - J_{yz}pr &= M_{aero,z}. \label{eq:rot_z}
\end{align}

\subsection{Quaternion Kinematics}
\label{sec:quaternion}

The attitude of the aircraft is represented by the unit quaternion $\boldsymbol{\zeta} = \{\zeta_0, \zeta_1, \zeta_2, \zeta_3\}^T$, where $\zeta_0$ is the scalar part and $\{\zeta_1, \zeta_2, \zeta_3\}$ is the vector part. The quaternion parametrisation avoids the gimbal lock singularity inherent in Euler angle representations, which is important for aircraft that may undergo large attitude changes (e.g., during aggressive manoeuvres or gust encounters at high dihedral).

The quaternion kinematic equation relates the time derivative of the quaternion to the body-frame angular velocity:
\begin{equation}
\dot{\boldsymbol{\zeta}} = \frac{1}{2}\boldsymbol{\Omega}(\boldsymbol{\omega})\,\boldsymbol{\zeta},
\label{eq:quaternion_kin}
\end{equation}
where $\boldsymbol{\Omega}(\boldsymbol{\omega})$ is the $4 \times 4$ skew-symmetric matrix:
\begin{equation}
\boldsymbol{\Omega}(\boldsymbol{\omega}) = \begin{bmatrix}
0 & -p & -q & -r \\
p & 0 & r & -q \\
q & -r & 0 & p \\
r & q & -p & 0
\end{bmatrix}.
\label{eq:omega_matrix}
\end{equation}

In expanded form, the four quaternion differential equations are:
\begin{align}
\dot{\zeta}_0 &= -\frac{1}{2}(p\,\zeta_1 + q\,\zeta_2 + r\,\zeta_3), \label{eq:quat_0} \\
\dot{\zeta}_1 &= \frac{1}{2}(p\,\zeta_0 + r\,\zeta_2 - q\,\zeta_3), \label{eq:quat_1} \\
\dot{\zeta}_2 &= \frac{1}{2}(q\,\zeta_0 - r\,\zeta_1 + p\,\zeta_3), \label{eq:quat_2} \\
\dot{\zeta}_3 &= \frac{1}{2}(r\,\zeta_0 + q\,\zeta_1 - p\,\zeta_2). \label{eq:quat_3}
\end{align}

The unit quaternion constraint $\|\boldsymbol{\zeta}\| = 1$ is enforced at each time step by normalisation:
\begin{equation}
\boldsymbol{\zeta} \leftarrow \frac{\boldsymbol{\zeta}}{\|\boldsymbol{\zeta}\|}.
\label{eq:quat_normalise}
\end{equation}

The rotation matrix from the body frame to the global (inertial) frame is obtained from the quaternion components as:
\begin{equation}
\mathbf{R}_\zeta = \begin{bmatrix}
1 - 2(\zeta_2^2 + \zeta_3^2) & 2(\zeta_1\zeta_2 - \zeta_0\zeta_3) & 2(\zeta_1\zeta_3 + \zeta_0\zeta_2) \\
2(\zeta_1\zeta_2 + \zeta_0\zeta_3) & 1 - 2(\zeta_1^2 + \zeta_3^2) & 2(\zeta_2\zeta_3 - \zeta_0\zeta_1) \\
2(\zeta_1\zeta_3 - \zeta_0\zeta_2) & 2(\zeta_2\zeta_3 + \zeta_0\zeta_1) & 1 - 2(\zeta_1^2 + \zeta_2^2)
\end{bmatrix}.
\label{eq:rotation_matrix}
\end{equation}

The inverse rotation (from global to body frame) is simply $\mathbf{R}_\zeta^T$, since $\mathbf{R}_\zeta \in SO(3)$.

\subsection{Position Integration}
\label{sec:position}

The position of the centre of mass in the global frame is obtained by integrating the velocity:
\begin{equation}
\dot{\mathbf{r}}_{cm} = \mathbf{R}_\zeta \, \mathbf{v},
\label{eq:position_integration}
\end{equation}
where $\mathbf{r}_{cm} = \{X, Y, Z\}^T$ is the position in the global frame and $\mathbf{v}$ is the body-frame velocity. This adds three states to the rigid-body subsystem, giving a total of 13 rigid-body states: 3 translational velocities $(u, v, w)$, 3 angular velocities $(p, q, r)$, 4 quaternion components $(\zeta_0, \zeta_1, \zeta_2, \zeta_3)$, and 3 positions $(X, Y, Z)$.

\subsection{Flexible-Body Corrections}
\label{sec:flex_corrections}

For a flexible aircraft, the centre of mass shifts as the structure deforms. The instantaneous centre of mass position in the body frame is
\begin{equation}
\mathbf{r}_{cm,B} = \frac{1}{m}\sum_{i=1}^{N_{nodes}} m_i \left(\mathbf{r}_{0,i} + \mathbf{u}_i\right),
\label{eq:com_shift}
\end{equation}
where $\mathbf{r}_{0,i}$ is the undeformed position of node $i$ and $\mathbf{u}_i$ is its elastic displacement. This shift couples the structural and flight dynamic subsystems: changes in structural deformation alter the centre of mass, which affects the moment arms of all forces and thus the rotational dynamics. Additionally, the inertia tensor $\mathbf{J}$ must be updated at each time step (or iteration) to reflect the current deformed shape.

\section{Coupled System Assembly}
\label{sec:coupling}

This section describes the assembly of the structural, aerodynamic, and flight dynamic subsystems into a single coupled system amenable to time-domain simulation, linearisation, and control design.

\subsection{State Vector Definition}
\label{sec:state_vector}

The complete state vector for the coupled system is
\begin{equation}
\mathbf{w} = \left\{ \mathbf{w}_f^T, \;\; \mathbf{w}_s^T, \;\; \dot{\mathbf{w}}_s^T, \;\; \mathbf{w}_r^T \right\}^T \in \mathbb{R}^n,
\label{eq:state_vector}
\end{equation}
where $\mathbf{w}_f \in \mathbb{R}^{n_f}$ collects the augmented aerodynamic states (Wagner + K\"ussner + flap, if any), with dimension $n_f = 4 N_s + 2 N_f$ (2 Wagner + 2 K\"ussner per strip, plus 2 per flapped strip); $\mathbf{w}_s \in \mathbb{R}^{n_s}$ contains the structural displacement DOFs (nodal translations and rotations), with $n_s = 6 \times N_{nodes}$; $\dot{\mathbf{w}}_s \in \mathbb{R}^{n_s}$ contains the structural velocity DOFs; and $\mathbf{w}_r \in \mathbb{R}^{13}$ contains the rigid-body states (3 translational velocities, 3 angular velocities, 4 quaternion components, 3 positions).

The total system dimension is $n = n_f + 2 n_s + 13$.

For a typical HALE aircraft model with 100 beam elements (101 nodes) and 50 aerodynamic strips (5 with flaps), the system dimension is:
\begin{equation}
n = (4 \times 50 + 2 \times 5) + 2 \times (6 \times 101) + 13 = 210 + 1212 + 13 = 1{,}435.
\end{equation}

\subsection{First-Order State-Space Form}
\label{sec:state_space}

The second-order structural equations, Eq.~\eqref{eq:eom}, are converted to first-order form by introducing the structural velocities as additional states. The complete coupled system is then written as
\begin{equation}
\dot{\mathbf{w}} = \mathbf{f}(\mathbf{w}, \mathbf{u}_c, \mathbf{u}_d),
\label{eq:nonlinear_ss}
\end{equation}
where $\mathbf{u}_c$ is the control input vector (e.g., flap deflections, thrust) and $\mathbf{u}_d$ is the disturbance (gust) input vector. The function $\mathbf{f}$ encapsulates the full nonlinear dynamics.

Linearisation about an equilibrium state $\mathbf{w}_0$ (obtained from a trim solution) gives the linear state-space system:
\begin{equation}
\Delta\dot{\mathbf{w}} = \mathbf{A}\,\Delta\mathbf{w} + \mathbf{B}_c\,\Delta\mathbf{u}_c + \mathbf{B}_g\,\mathbf{u}_d,
\label{eq:linear_ss}
\end{equation}
where $\mathbf{A} = \partial\mathbf{f}/\partial\mathbf{w}\big|_{\mathbf{w}_0}$ is the Jacobian matrix, $\mathbf{B}_c = \partial\mathbf{f}/\partial\mathbf{u}_c\big|_{\mathbf{w}_0}$ is the control input matrix, and $\mathbf{B}_g = \partial\mathbf{f}/\partial\mathbf{u}_d\big|_{\mathbf{w}_0}$ is the gust input matrix. The symbol $\Delta$ denotes perturbation from the equilibrium.

For nonlinear time-domain simulation, the full nonlinear system, Eq.~\eqref{eq:nonlinear_ss}, is integrated directly. The linearised system, Eq.~\eqref{eq:linear_ss}, is used for stability analysis (eigenvalue computation), model order reduction~\citep{DaRonch2013control, Tantaroudas2015scitech}, and linear control design~\citep{Tantaroudas2014aviation}.

\subsection{Coordinate Transformations in the Coupled System}
\label{sec:coord_transform}

The coupling between the subsystems occurs through coordinate transformations that link the aerodynamic, structural, and flight dynamic frames. The key transformations are:

\paragraph{Effective velocity at each aerodynamic strip.} The freestream velocity seen by each strip depends on the rigid-body motion, the elastic deformation, and the gust velocity. In the body frame:
\begin{equation}
\mathbf{U}_{eff,j} = \mathbf{R}_\zeta^T \mathbf{U}_\infty - \boldsymbol{\omega} \times \mathbf{r}_j - \dot{\mathbf{u}}_j + \mathbf{R}_\zeta^T \mathbf{w}_{g,j},
\label{eq:U_eff}
\end{equation}
where $\mathbf{U}_\infty$ is the freestream velocity in the global frame, $\mathbf{r}_j = \mathbf{r}_{0,j} + \mathbf{u}_j$ is the deformed position of strip $j$ relative to the centre of mass, $\dot{\mathbf{u}}_j$ is the structural velocity at strip $j$, and $\mathbf{w}_{g,j}$ is the gust velocity at strip $j$ in the global frame. This equation is the primary coupling mechanism: the aerodynamic loads depend on the rigid-body state (through $\mathbf{R}_\zeta$ and $\boldsymbol{\omega}$), the structural state (through $\mathbf{r}_j$ and $\dot{\mathbf{u}}_j$), and the gust input (through $\mathbf{w}_{g,j}$).

\paragraph{Local angle of attack and sideslip.} The local effective velocity at strip $j$ is decomposed into components to obtain the local angle of attack and sideslip in the aerodynamic frame:
\begin{equation}
\alpha_j = \arctan\left(\frac{U_{eff,j}^z}{U_{eff,j}^x}\right), \qquad \beta_j = \arcsin\left(\frac{U_{eff,j}^y}{\|\mathbf{U}_{eff,j}\|}\right),
\label{eq:local_aoa}
\end{equation}
where the superscripts denote components in the local beam frame after appropriate transformation. For small angles, the linearised form $\alpha_j \approx U_{eff,j}^z / U_{eff,j}^x$ is used.

\paragraph{Aerodynamic-to-structural force transformation.} The aerodynamic forces computed in the aerodynamic frame are transformed to the global frame via:
\begin{equation}
\mathbf{f}_{G,j} = \mathbf{C}_{BG,j}^T \, \mathbf{R}_{c,j} \, \mathbf{f}_{A,j},
\label{eq:aero_to_global}
\end{equation}
where $\mathbf{C}_{BG,j}$ is the beam-to-global rotation at strip $j$ (including elastic deformation) and $\mathbf{R}_{c,j}$ is the aerodynamic-to-beam rotation. These forces are then assembled into the global force vector for the structural equations of motion.

\subsection{Jacobian Block Structure}
\label{sec:jacobian}

The Jacobian matrix $\mathbf{A}$ of the linearised coupled system has a characteristic block structure that reflects the physical coupling between the three subsystems~\citep{Hesse2014, Tantaroudas2017bookchapter}:
\begin{equation}
\mathbf{A} = \begin{bmatrix}
\mathbf{A}_{ff} & \mathbf{A}_{fs} & \mathbf{A}_{f\dot{s}} & \mathbf{A}_{fr} \\[4pt]
\mathbf{0} & \mathbf{0} & \mathbf{I} & \mathbf{0} \\[4pt]
\mathbf{A}_{\dot{s}f} & \mathbf{A}_{\dot{s}s} & \mathbf{A}_{\dot{s}\dot{s}} & \mathbf{A}_{\dot{s}r} \\[4pt]
\mathbf{A}_{rf} & \mathbf{A}_{rs} & \mathbf{A}_{r\dot{s}} & \mathbf{A}_{rr}
\end{bmatrix},
\label{eq:jacobian_block}
\end{equation}
where the second row implements the identity $\dot{\mathbf{w}}_s = \dot{\mathbf{w}}_s$ (the kinematic relationship between displacement and velocity states). The physical meaning of each block is:

The block $\mathbf{A}_{ff} \in \mathbb{R}^{n_f \times n_f}$ governs the aerodynamic state dynamics and is a diagonal (or block-diagonal) matrix containing the exponential decay rates $-\varepsilon_k U/b$ for the Wagner and K\"ussner states; this block is always stable. The blocks $\mathbf{A}_{fs} \in \mathbb{R}^{n_f \times n_s}$ and $\mathbf{A}_{f\dot{s}} \in \mathbb{R}^{n_f \times n_s}$ represent the coupling from structural displacements and velocities to the aerodynamic states, arising because the effective angle of attack depends on the structural motion, while $\mathbf{A}_{fr} \in \mathbb{R}^{n_f \times 13}$ captures the coupling from rigid-body states to aerodynamic states through changes in body velocity and angular velocity. The block $\mathbf{A}_{\dot{s}f} \in \mathbb{R}^{n_s \times n_f}$ represents the structural forcing from the circulatory lift history stored in the augmented states, proportional to $\mathbf{M}^{-1}\mathbf{F}_{aero}^{aug}$. The block $\mathbf{A}_{\dot{s}s} \in \mathbb{R}^{n_s \times n_s}$ contains the combined structural and aerodynamic stiffness $-\mathbf{M}^{-1}(\mathbf{K}_s - \mathbf{K}_a)$, whose competition determines static divergence, while $\mathbf{A}_{\dot{s}\dot{s}} \in \mathbb{R}^{n_s \times n_s}$ contains the combined structural and aerodynamic damping $-\mathbf{M}^{-1}(\mathbf{C}_s - \mathbf{C}_a)$, whose sign determines flutter stability. The block $\mathbf{A}_{\dot{s}r} \in \mathbb{R}^{n_s \times 13}$ captures the structural coupling to rigid-body states, including the dependence of aerodynamic forces on body velocity and angular velocity, as well as centrifugal and Coriolis terms. The blocks $\mathbf{A}_{rf}$, $\mathbf{A}_{rs}$, $\mathbf{A}_{r\dot{s}} \in \mathbb{R}^{13 \times (\cdot)}$ represent the rigid-body forcing from aerodynamic and structural states, as the total aerodynamic force and moment depend on the circulatory lift, the structural displacements, and the structural velocities. Finally, $\mathbf{A}_{rr} \in \mathbb{R}^{13 \times 13}$ contains the rigid-body self-dynamics, including the linearised Newton--Euler equations and the quaternion kinematic equations.

\subsection{Gust Input Matrix}
\label{sec:gust_input}

The gust input matrix $\mathbf{B}_g$ describes how the gust velocity at each strip enters the coupled system. For a vertical gust $w_g$ propagating along the flight direction, the gust input affects the aerodynamic augmented states through the K\"ussner memory (contributing to the $\mathbf{w}_f$ dynamics), the aerodynamic forces on the structure (contributing to the $\dot{\mathbf{w}}_s$ dynamics through non-circulatory gust loads), and the rigid-body forces and moments through the integrated gust-induced loads.

The gust input matrix has the structure:
\begin{equation}
\mathbf{B}_g = \begin{bmatrix} \mathbf{B}_{g,f} \\ \mathbf{0} \\ \mathbf{B}_{g,\dot{s}} \\ \mathbf{B}_{g,r} \end{bmatrix},
\label{eq:gust_input_matrix}
\end{equation}
where $\mathbf{B}_{g,f}$ drives the K\"ussner augmented states, $\mathbf{B}_{g,\dot{s}}$ contributes non-circulatory gust loads to the structural acceleration, and $\mathbf{B}_{g,r}$ contributes gust-induced forces and moments to the rigid-body dynamics.

\subsection{Control Input Matrix}
\label{sec:control_input}

The control input matrix $\mathbf{B}_c$ describes how control surface deflections and thrust variations enter the system. For trailing-edge flaps, the control input affects the aerodynamic augmented states for the flap (Wagner-type memory of flap deflection), the structural forces through flap-induced lift, moment, and hinge moment, and the rigid-body forces and moments through the integrated flap loads.

The control input matrix has a similar block structure to $\mathbf{B}_g$ and is derived by differentiating the nonlinear equations with respect to the control inputs at the trim condition.

\section{Gust Models}
\label{sec:gust}

\begin{figure}[htbp]
\centering
\includegraphics[width=0.65\textwidth]{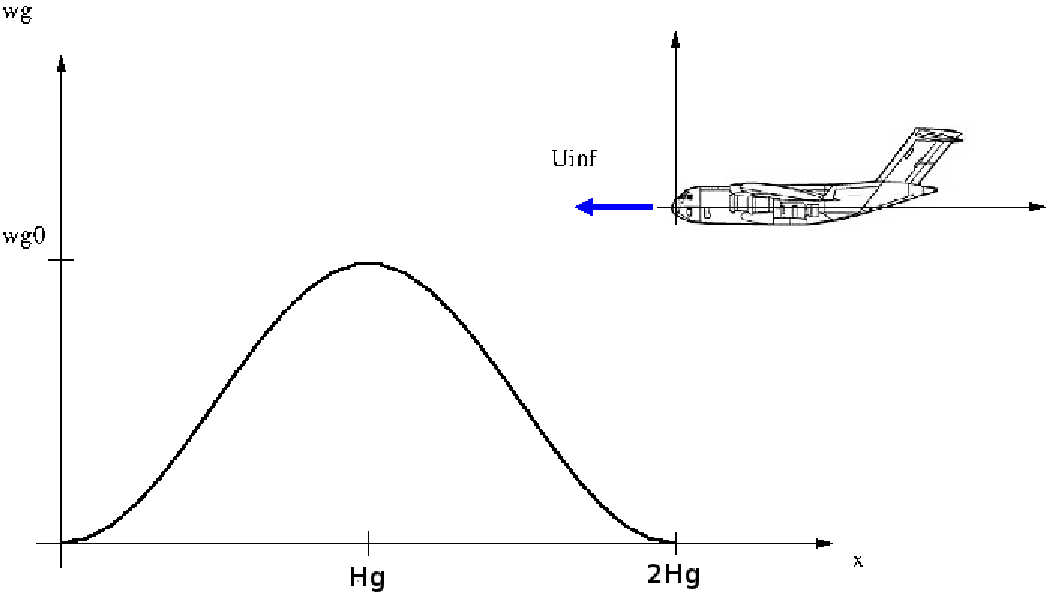}
\caption{1-minus-cosine discrete gust profile. The gust gradient distance $H_g$ defines the spatial extent of the gust, and $W_0$ is the peak gust velocity.}
\label{fig:omc_gust}
\end{figure}

Two types of atmospheric gust models are used in the present framework: the discrete ``1-minus-cosine'' gust for certification analysis (Figure~\ref{fig:omc_gust}) and the Von K\'arm\'an continuous turbulence spectrum for stochastic analysis.

\subsection{Discrete Gust: 1-Minus-Cosine Profile}
\label{sec:discrete_gust}

The 1-minus-cosine gust is prescribed by the certification regulations CS-25 and FAR-25 for discrete gust encounter analysis. The vertical gust velocity is defined as
\begin{equation}
w_g(x) = \begin{cases}
\displaystyle\frac{W_0}{2}\left(1 - \cos\left(\frac{\pi x}{H_g}\right)\right), & 0 \leq x \leq 2H_g, \\[8pt]
0, & \text{otherwise},
\end{cases}
\label{eq:1mc_gust}
\end{equation}
where $W_0$ is the peak gust velocity (design gust velocity), $H_g$ is the gust gradient distance (the distance from zero gust velocity to the peak, equal to half the gust wavelength), and $x = U(t - t_0)$ is the distance travelled into the gust from the gust front at time $t_0$.

The design gust velocity depends on the gust gradient distance according to CS-25.341:
\begin{equation}
W_0 = W_{ref} F_g \left(\frac{H_g}{107}\right)^{1/6},
\label{eq:design_gust}
\end{equation}
where $W_{ref}$ is the reference gust velocity (which depends on altitude and flight speed), $F_g$ is the flight profile alleviation factor, and $H_g$ is expressed in feet (ranging from 30 ft to 350 ft). Multiple gust gradient distances must be analysed to find the critical (worst-case) response.

The temporal gust velocity at the aircraft reference point (the nose or leading edge of the wing root) is
\begin{equation}
w_g(t) = \frac{W_0}{2}\left(1 - \cos\left(\frac{\pi U (t-t_0)}{H_g}\right)\right), \qquad t_0 \leq t \leq t_0 + \frac{2H_g}{U}.
\label{eq:1mc_temporal}
\end{equation}

The time derivative, needed for the K\"ussner augmented state equations (Eq.~\eqref{eq:kussner_ode}), is
\begin{equation}
\dot{w}_g(t) = \frac{\pi U W_0}{2H_g}\sin\left(\frac{\pi U(t-t_0)}{H_g}\right).
\label{eq:1mc_derivative}
\end{equation}

\subsection{Continuous Turbulence: Von K\'arm\'an Spectrum}
\label{sec:von_karman}

For continuous turbulence analysis, the Von K\'arm\'an power spectral density (PSD) model provides a physically-based representation of atmospheric turbulence. The vertical component of the turbulence spectrum is~\citep{Dowell2004}:
\begin{equation}
\Phi_{ww}(\Omega) = \sigma_w^2 \frac{L_w}{\pi} \frac{1 + \frac{8}{3}\left(1.339\,L_w\Omega\right)^2}{\left(1 + \left(1.339\,L_w\Omega\right)^2\right)^{11/6}},
\label{eq:von_karman_psd}
\end{equation}
where $\sigma_w$ is the root-mean-square (RMS) turbulence intensity (m/s), $L_w$ is the turbulence scale length (m), and $\Omega = 2\pi f / U$ is the spatial frequency (rad/m). The turbulence scale length is altitude-dependent: $L_w = 2500$ ft (762 m) above 2000 ft altitude.

For time-domain simulation, the Von K\'arm\'an spectrum is approximated by a rational transfer function (coloured noise filter) that can be driven by white noise $\eta(t)$:
\begin{equation}
w_g(s) = \sigma_w \sqrt{\frac{2L_w}{\pi U}} \cdot \frac{1 + 2.618 \frac{L_w}{U}s + 0.1767 \left(\frac{L_w}{U}\right)^2 s^2}{\left(1 + 2.618 \frac{L_w}{U}s\right)\left(1 + 0.1767 \frac{L_w}{U}s\right)^2} \cdot \eta(s),
\label{eq:von_karman_filter}
\end{equation}
where $s$ is the Laplace variable. This transfer function is implemented as a state-space system with additional internal states that are appended to the gust input processing.

The lateral and longitudinal turbulence components have similar spectral forms with different scale lengths, but the vertical component is typically the most critical for flexible aircraft gust response.

\subsection{Gust Penetration Effect}
\label{sec:gust_penetration}

For a finite-span aircraft, the gust does not arrive simultaneously at all spanwise strips. As the aircraft flies into the gust field, the strips at different chordwise positions encounter the gust at different times. The gust velocity at strip $j$, located at a chordwise distance $x_j$ from the aircraft reference point, is delayed:
\begin{equation}
w_{g,j}(t) = w_g\left(t - \frac{x_j}{U}\right),
\label{eq:gust_penetration}
\end{equation}
where $x_j$ is the distance of strip $j$ aft of the reference point along the flight direction. For a swept wing, the strips at the tips are further aft than the root, so they encounter the gust later~\citep{DaRonch2013gust}.

This spanwise delay is incorporated into the gust input matrix $\mathbf{B}_g$ by assigning different time delays to each strip. In the time-domain simulation, this is implemented by maintaining a gust velocity history buffer and interpolating the gust velocity for each strip at its delayed time.

The gust penetration effect is particularly important for short-wavelength gusts (small $H_g$) and for aircraft with large wing spans and/or significant wing sweep, where the temporal delay between root and tip can be a significant fraction of the gust duration.

\section{Time Integration}
\label{sec:integration}

The coupled first-order system, Eq.~\eqref{eq:nonlinear_ss}, is integrated in the time domain using different schemes for the different subsystems, chosen to match the numerical characteristics of each.

\subsection[Newmark-beta Scheme for Structural Dynamics]{Newmark-$\beta$ Scheme for Structural Dynamics}
\label{sec:newmark}

The structural degrees of freedom are integrated using the Newmark-$\beta$ implicit time integration scheme~\citep{Newmark1959}. The displacement and velocity updates from time step $n$ to $n+1$ are:
\begin{align}
\mathbf{w}_{s,n+1} &= \mathbf{w}_{s,n} + \Delta t \, \dot{\mathbf{w}}_{s,n} + \frac{\Delta t^2}{2}\left[(1-2\beta_N)\ddot{\mathbf{w}}_{s,n} + 2\beta_N\ddot{\mathbf{w}}_{s,n+1}\right], \label{eq:newmark_disp} \\
\dot{\mathbf{w}}_{s,n+1} &= \dot{\mathbf{w}}_{s,n} + \Delta t\left[(1-\gamma_N)\ddot{\mathbf{w}}_{s,n} + \gamma_N\ddot{\mathbf{w}}_{s,n+1}\right], \label{eq:newmark_vel}
\end{align}
where $\Delta t$ is the time step, $\ddot{\mathbf{w}}_{s,n+1}$ is the unknown acceleration at the new time level, and $\beta_N$ and $\gamma_N$ are the Newmark parameters.

The parameters are chosen as $\gamma_N = 0.51$ and $\beta_N = 0.255025$ (satisfying the stability condition $\beta_N \geq \gamma_N/2$ and $\gamma_N \geq 1/2$). This choice introduces slight numerical dissipation ($\gamma_N > 1/2$) that damps high-frequency numerical oscillations without significantly affecting the low-frequency structural modes of interest. The trapezoidal rule ($\gamma_N = 1/2$, $\beta_N = 1/4$) is energy-conserving but can exhibit high-frequency ringing in the presence of numerical noise.

The unknown acceleration $\ddot{\mathbf{w}}_{s,n+1}$ is obtained by solving the structural equations of motion at time $t_{n+1}$:
\begin{equation}
\mathbf{M}\,\ddot{\mathbf{w}}_{s,n+1} = \mathbf{R}_F(t_{n+1}, \mathbf{w}_{s,n+1}, \dot{\mathbf{w}}_{s,n+1}) - \mathbf{Q}_{gyr}\,\dot{\mathbf{w}}_{n+1} - \mathbf{Q}_{stiff}\,\mathbf{w}_{n+1}.
\label{eq:newmark_residual}
\end{equation}

Since $\mathbf{w}_{s,n+1}$ and $\dot{\mathbf{w}}_{s,n+1}$ depend on the unknown $\ddot{\mathbf{w}}_{s,n+1}$ through the Newmark relations, the system is implicit and must be solved iteratively (typically using Newton--Raphson iteration) at each time step. From Eqs.~\eqref{eq:newmark_disp}--\eqref{eq:newmark_vel}, the effective relationships are:
\begin{align}
\ddot{\mathbf{w}}_{s,n+1} &= \frac{1}{\beta_N \Delta t^2}\left(\mathbf{w}_{s,n+1} - \mathbf{w}_{s,n} - \Delta t\,\dot{\mathbf{w}}_{s,n}\right) - \frac{1-2\beta_N}{2\beta_N}\ddot{\mathbf{w}}_{s,n}, \label{eq:newmark_acc} \\
\dot{\mathbf{w}}_{s,n+1} &= \frac{\gamma_N}{\beta_N \Delta t}\left(\mathbf{w}_{s,n+1} - \mathbf{w}_{s,n}\right) + \left(1 - \frac{\gamma_N}{\beta_N}\right)\dot{\mathbf{w}}_{s,n} + \Delta t\left(1 - \frac{\gamma_N}{2\beta_N}\right)\ddot{\mathbf{w}}_{s,n}. \label{eq:newmark_vel2}
\end{align}

The effective stiffness matrix for the Newton--Raphson iteration is
\begin{equation}
\mathbf{K}_{eff} = \frac{1}{\beta_N \Delta t^2}\mathbf{M} + \frac{\gamma_N}{\beta_N \Delta t}\mathbf{C} + \mathbf{K}_{tan},
\label{eq:effective_stiffness}
\end{equation}
where $\mathbf{C}$ includes both structural damping and aerodynamic damping terms.

\subsection{Integration of Aerodynamic States}
\label{sec:aero_integration}

The augmented aerodynamic states (Wagner and K\"ussner) satisfy first-order linear ODEs of the form $\dot{\lambda} = -a\lambda + b(t)$, where $a > 0$ is the exponential decay rate. These are integrated using the backward Euler (implicit Euler) scheme:
\begin{equation}
\lambda_{n+1} = \frac{\lambda_n + \Delta t \, b(t_{n+1})}{1 + a\,\Delta t},
\label{eq:backward_euler}
\end{equation}
which is unconditionally stable and perfectly suited for the exponentially decaying augmented states. The backward Euler scheme introduces numerical dissipation, but this is acceptable for the augmented states since they represent wake memory effects that decay physically.

\subsection{Integration of Rigid-Body States}
\label{sec:rb_integration}

The rigid-body translational and rotational velocity equations are integrated using the same time step $\Delta t$. The quaternion kinematic equations are integrated explicitly (forward Euler or second-order Runge--Kutta), and the quaternion is normalised at each time step:
\begin{equation}
\boldsymbol{\zeta}_{n+1} = \frac{\boldsymbol{\zeta}_n + \frac{\Delta t}{2}\boldsymbol{\Omega}(\boldsymbol{\omega}_n)\boldsymbol{\zeta}_n}{\left\|\boldsymbol{\zeta}_n + \frac{\Delta t}{2}\boldsymbol{\Omega}(\boldsymbol{\omega}_n)\boldsymbol{\zeta}_n\right\|}.
\label{eq:quat_integration}
\end{equation}

The position is integrated using the trapezoidal rule:
\begin{equation}
\mathbf{r}_{cm,n+1} = \mathbf{r}_{cm,n} + \frac{\Delta t}{2}\left(\mathbf{R}_{\zeta,n}\mathbf{v}_n + \mathbf{R}_{\zeta,n+1}\mathbf{v}_{n+1}\right).
\label{eq:position_trapezoidal}
\end{equation}

\subsection{Time Step Selection}
\label{sec:time_step}

The time step $\Delta t$ must be small enough to resolve the highest-frequency dynamics of interest. For aeroelastic analysis, the critical frequencies are typically the structural natural frequencies and the aerodynamic reduced frequencies. A guideline is $\Delta t \leq T_{min}/20$, where $T_{min}$ is the period of the highest mode of interest. For gust response analysis, the time step must also resolve the gust temporal profile: $\Delta t \leq H_g/(20U)$ for the 1-minus-cosine gust. Typical time steps range from $10^{-3}$ to $10^{-2}$ seconds.

\section{Verification and Results}
\label{sec:verification}

The coupled aeroelastic-flight dynamic framework is verified against two established benchmarks: the Patil HALE aircraft and a very flexible flying-wing configuration.

\subsection{HALE Aircraft Configuration}
\label{sec:hale_config}

\begin{figure}[htbp]
\centering
\includegraphics[width=0.8\textwidth]{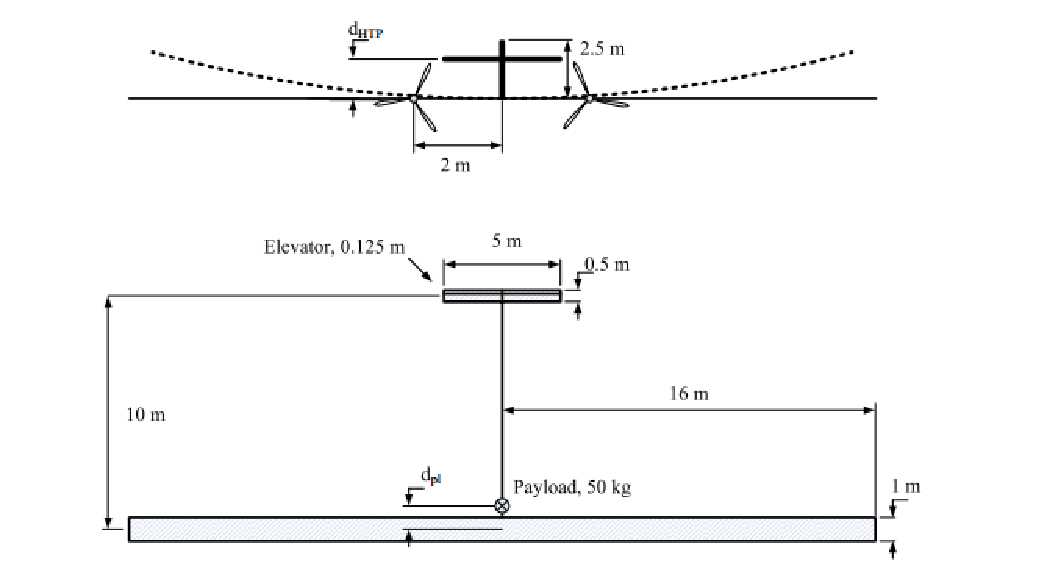}
\caption{HALE aircraft configuration showing the wing, fuselage, horizontal tail, vertical tail, and propeller locations. The wing has a 32-m span, 1-m chord, and is clamped to the fuselage at mid-span.}
\label{fig:hale_geometry}
\end{figure}

The HALE aircraft configuration of \citet{Patil2001} is widely used as a benchmark for flexible aircraft analysis tools. It consists of a high-aspect-ratio wing with a rigid fuselage, horizontal and vertical tail surfaces, and two propellers (Figure~\ref{fig:hale_geometry}).

The key parameters of the HALE aircraft are as follows. The wing has a span of $b_w = 32$ m ($2 \times 16$ m semi-spans) with a chord of $c = 1$ m, elastic axis at 50\% chord, and centre of gravity at 50\% chord. The stiffness properties are $EI_2 = 2 \times 10^4$ Nm$^2$ (flapwise bending), $EI_3 = 5 \times 10^6$ Nm$^2$ (edgewise bending), and $GJ = 1 \times 10^4$ Nm$^2$ (torsion). The mass per unit span is $\mu = 0.75$ kg/m, with a total payload at the fuselage of $m_p = 50$ kg. The tail surfaces have a horizontal tail semi-span of 2.5 m and a vertical tail semi-span of 2.5 m, with proportional stiffness and mass properties. The flight condition is $\rho = 0.0889$ kg/m$^3$ (20 km altitude), with a speed range of 20--40 m/s.

The finite element model (Figure~\ref{fig:hale_fem}) uses 100 two-noded beam elements (20 per wing semi-span, plus tail and fuselage elements), yielding $N_{nodes} = 101$ and $n_s = 606$ displacement DOFs. With 50 aerodynamic strips (25 per wing semi-span) and 13 rigid-body states, the total system dimension is approximately 1,435 first-order states~\citep{Tantaroudas2015scitech, DaRonch2014scitech_flight}.

\subsection{Structural Natural Frequencies}
\label{sec:frequencies}

The structural natural frequencies of the clamped wing are compared with those reported by \citet{Patil2001} and \citet{Murua2012} in Table~\ref{tab:frequencies}. Excellent agreement is obtained for the first four modes. The frequencies are computed from the eigenvalues of the undamped structural system $\mathbf{K}_s \boldsymbol{\phi} = \omega^2 \mathbf{M}_{ss} \boldsymbol{\phi}$.

\begin{table}[htbp]
\centering
\caption{Structural natural frequencies of the HALE wing (clamped at root), in rad/s.}
\label{tab:frequencies}
\begin{tabular}{lccc}
\toprule
Mode & Present & Patil et al.~\citep{Patil2001} & Exact beam theory \\
\midrule
1st bending & 2.24 & 2.24 & 2.24 \\
2nd bending & 14.07 & 14.60 & 14.05 \\
1st torsion & 31.04 & 31.14 & 31.04 \\
1st in-plane bending & 31.71 & 31.73 & 31.71 \\
3rd bending & 39.52 & 44.01 & 39.35 \\
\bottomrule
\end{tabular}
\end{table}

The close agreement confirms the correct implementation of the structural dynamics model, including the mass and stiffness distributions.

\subsection{Flutter Speed}
\label{sec:flutter}

The flutter speed of the clamped HALE wing is determined from the eigenvalue analysis of the linearised aeroelastic system (without rigid-body DOFs). The system Jacobian is computed at increasing freestream velocities, and the flutter speed is identified as the velocity at which an eigenvalue crosses the imaginary axis (zero real part).

\begin{figure}[htbp]
\centering
\includegraphics[width=0.7\textwidth]{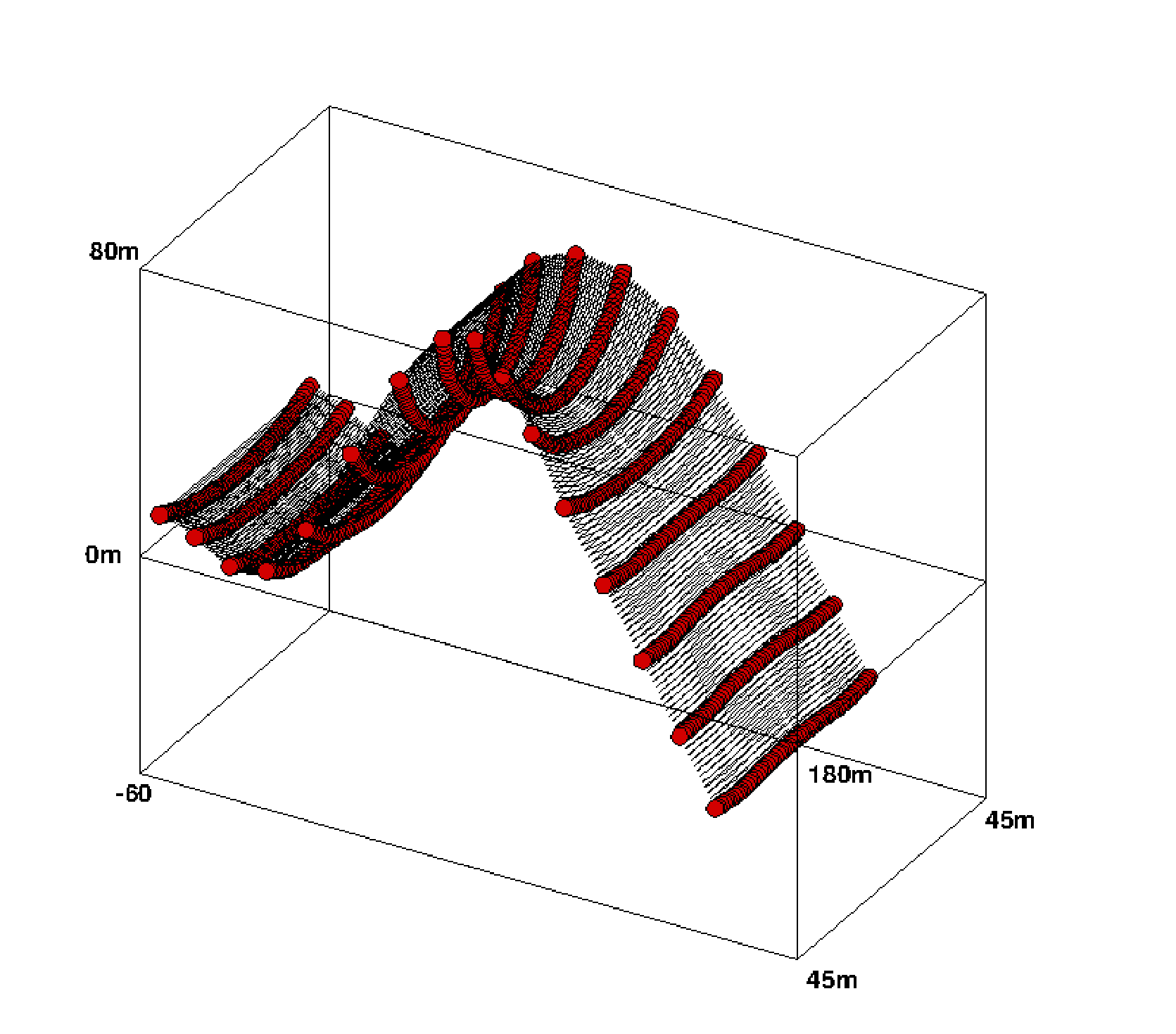}
\caption{Nonlinear flutter response of the HALE wing: three-dimensional visualisation of the wing deformation time history beyond the flutter boundary. Multiple time snapshots are superimposed, showing the growing oscillatory structural response characteristic of post-flutter behaviour. Red markers indicate the finite element node locations at successive time instants.}
\label{fig:flutter}
\end{figure}

The present framework predicts a flutter speed of 31.2 m/s, compared to 32.2 m/s reported by \citet{Patil2006} and 33.0 m/s by \citet{Murua2012}. The flutter mechanism involves the coalescence of the first torsion and second flapwise bending modes. Figure~\ref{fig:flutter} shows the nonlinear time-domain response of the HALE wing at a speed above the flutter boundary, where multiple time snapshots of the deformation are superimposed to visualise the growing oscillatory motion. The small differences between the predicted flutter speeds are attributed to differences in aerodynamic modelling: the present strip-theory model tends to predict slightly lower flutter speeds than UVLM~\citep{Murua2012} due to the neglect of three-dimensional aerodynamic effects at the wing tip, which reduce the effective lift curve slope and thus raise the flutter speed.

\subsection{Static Aeroelastic Deflections and Trim}
\label{sec:static}

The static aeroelastic trim solution is obtained by setting all time derivatives to zero in the coupled system, Eq.~\eqref{eq:nonlinear_ss}, and solving the resulting nonlinear algebraic system using Newton--Raphson iteration. The unknowns are the structural displacements, the trim angle of attack, and the thrust required for level flight.

\begin{figure}[htbp]
\centering
\includegraphics[width=0.7\textwidth]{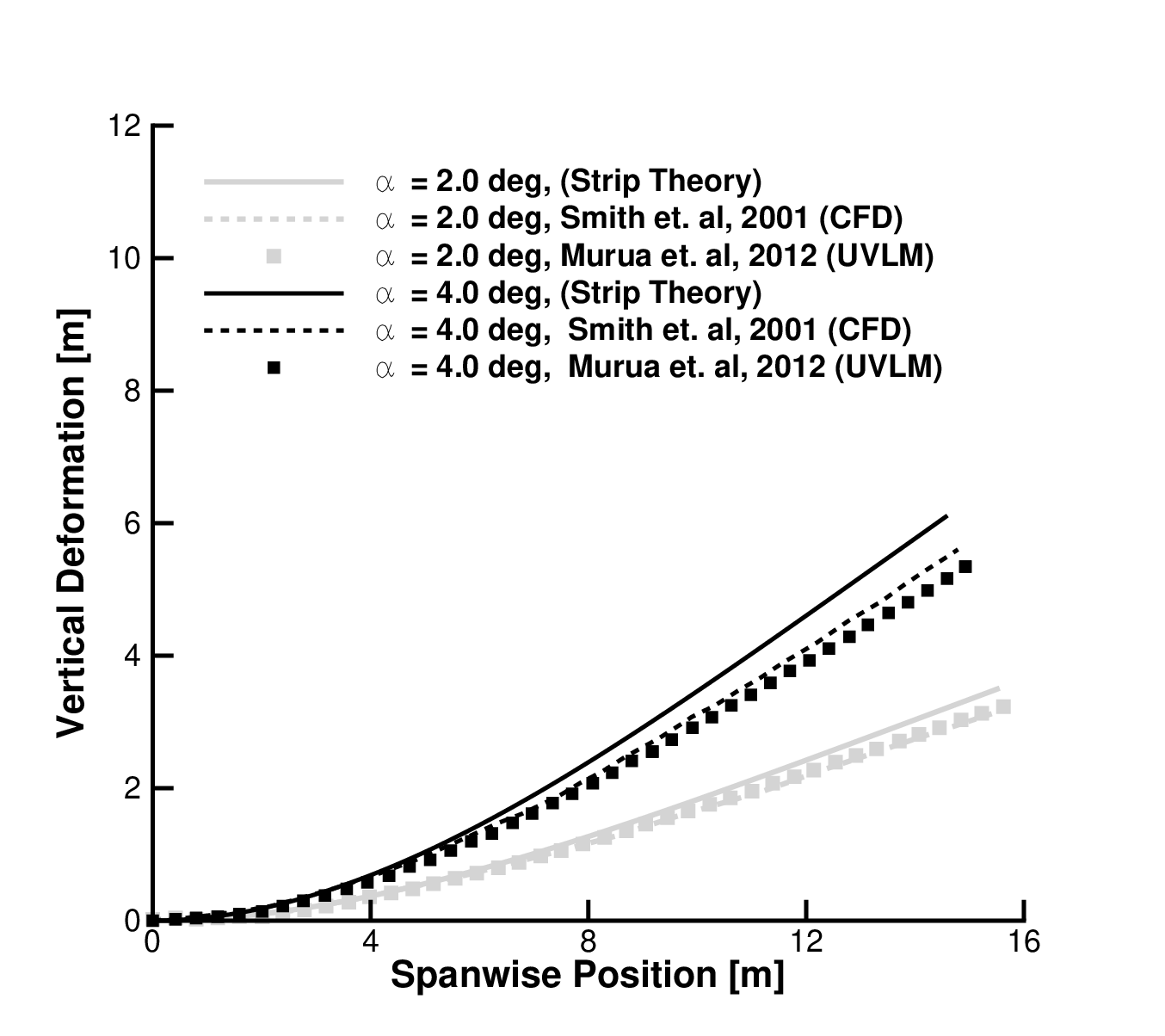}
\caption{Static aeroelastic deflection of the HALE wing semi-span at $U = 25$ m/s and 20 km altitude for two angles of attack ($\alpha = 2^\circ$ and $\alpha = 4^\circ$), comparing the present strip theory with CFD data of \citet{Smith2001} and UVLM results of \citet{Murua2012}.}
\label{fig:hale_static}
\end{figure}

Figure~\ref{fig:hale_static} shows the static aeroelastic deflection of the HALE wing at trim for two angles of attack ($\alpha = 2^\circ$ and $\alpha = 4^\circ$). The wing tip displacement is significant, confirming the need for geometrically nonlinear structural analysis. The present strip theory results are compared with the CFD data of \citet{Smith2001} and the UVLM results of \citet{Murua2012}, showing good agreement. Strip theory slightly overpredicts the deflection at higher angles of attack due to the neglect of three-dimensional aerodynamic effects and viscous corrections~\citep{DaRonch2014scitech_flight}.

\begin{figure}[htbp]
\centering
\includegraphics[width=0.7\textwidth]{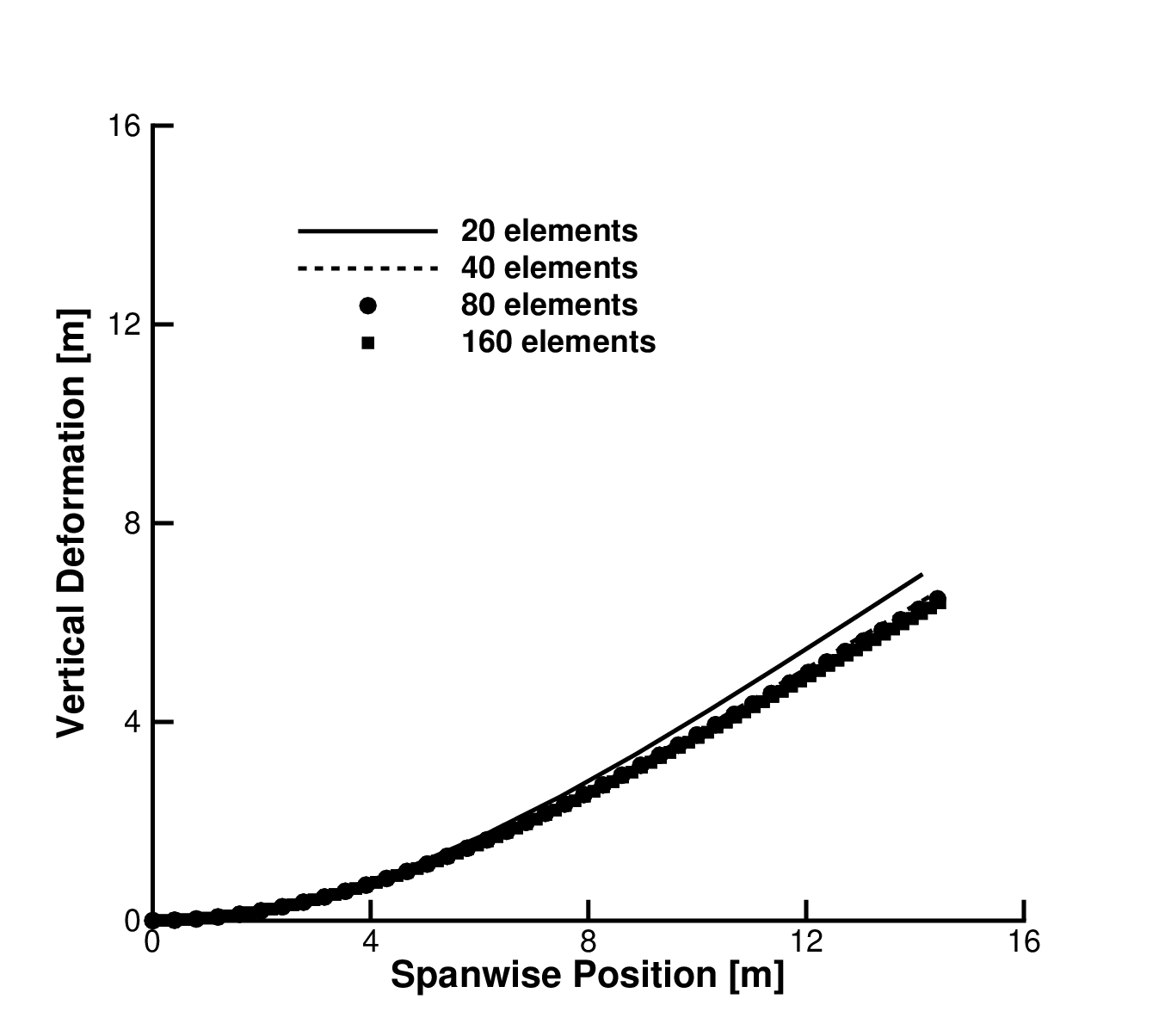}
\caption{Mesh convergence study for the static aeroelastic deflection of the HALE wing. The vertical deformation along the semi-span is computed using 20, 40, 80, and 160 two-noded beam elements. Convergence is achieved with 40 elements, confirming that the baseline mesh of 20 elements per semi-span provides adequate accuracy.}
\label{fig:hale_deformation}
\end{figure}

As shown in Figure~\ref{fig:hale_deformation}, the mesh convergence study confirms that the finite element discretisation with 20 elements per semi-span is sufficient for the static aeroelastic analysis, with the solution converging rapidly as the mesh is refined. The large wing deformation at trim creates significant dihedral, which has important consequences for flight dynamics~\citep{Patil2006, Tantaroudas2017bookchapter}: the local lift vectors tilt inward (toward the fuselage), reducing the total vertical lift component and requiring a higher trim angle of attack compared to the rigid case; the dihedral introduces lateral stability derivatives that are absent in the undeformed configuration; the effective moment arms of the aerodynamic forces change, altering the pitch, roll, and yaw stability; and the inertia tensor of the aircraft changes, modifying the natural frequencies of the rigid-body modes.

These coupled effects demonstrate why an integrated aeroelastic-flight dynamic analysis is essential for very flexible aircraft.

\subsection{Very Flexible Flying-Wing}
\label{sec:vfa}

A second verification case is a 32-m span very flexible flying-wing configuration (without tail surfaces), representative of solar-powered HALE platforms~\citep{Tantaroudas2015scitech}. The key parameters are: span 32 m with chord 1 m and elastic axis at 25\% chord; mass per unit span $\mu = 10$ kg/m; bending stiffness $EI_2 = 2.5 \times 10^4$ Nm$^2$ and torsional stiffness $GJ = 1.25 \times 10^4$ Nm$^2$; 10\% chord trailing-edge flaps on the outboard 50\% of each semi-span; and a finite element model of 80 two-noded beam elements (81 nodes, 486 structural displacement DOFs), yielding approximately 1,145 first-order states when augmented aerodynamic states, velocities, and rigid-body DOFs are included.

\begin{figure}[htbp]
\centering
\includegraphics[width=0.7\textwidth]{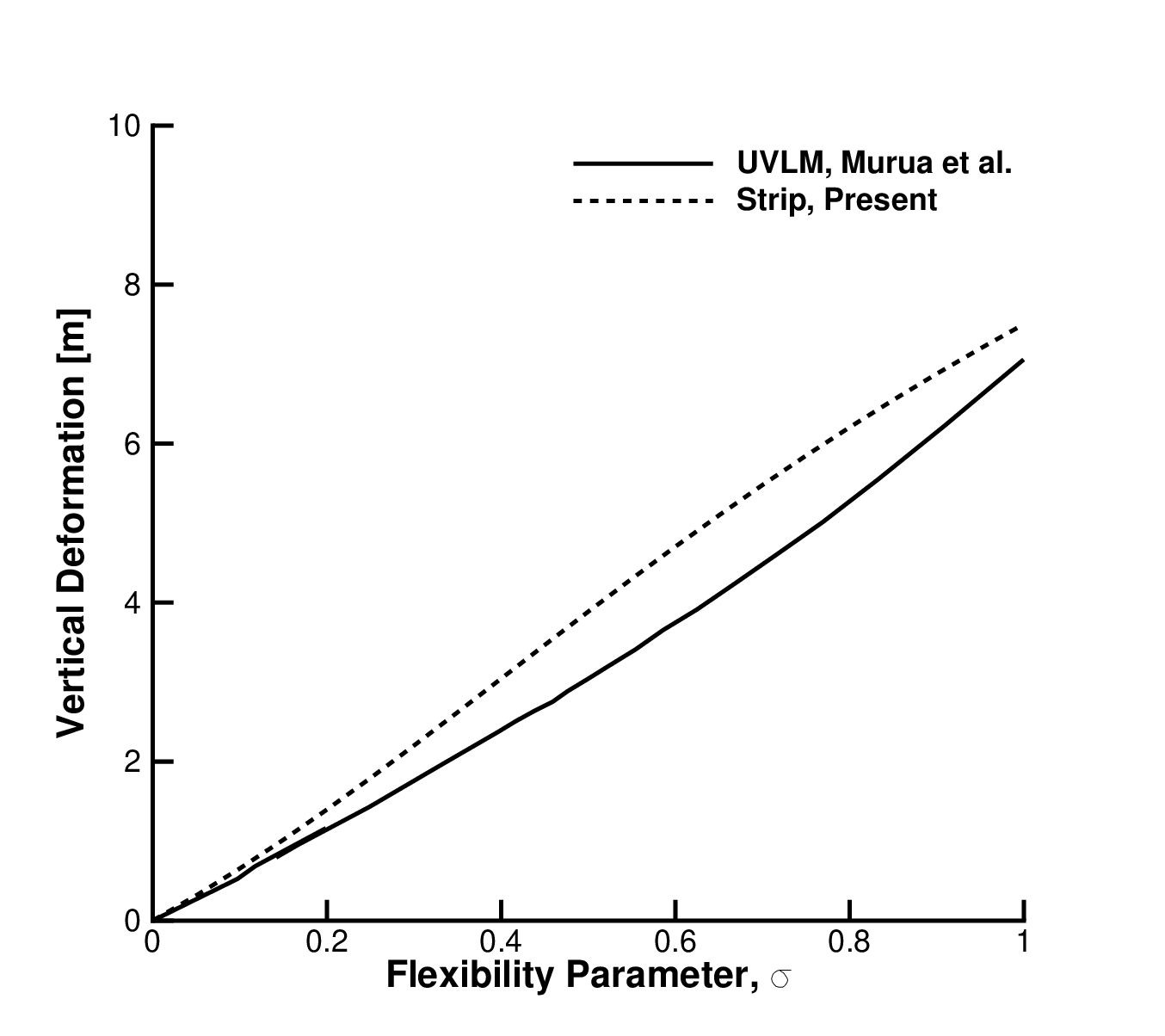}
\caption{Wing tip vertical deformation of the very flexible flying-wing as a function of the non-dimensional flexibility parameter $\sigma$ (ratio of actual-to-baseline stiffness inverse), comparing the present strip theory results with UVLM data of \citet{Murua2012}. Strip theory slightly overpredicts the deflection owing to the neglect of three-dimensional aerodynamic effects.}
\label{fig:vfa_static}
\end{figure}

Figure~\ref{fig:vfa_static} shows the wing tip vertical deformation of the VFA as a function of a non-dimensional flexibility parameter $\sigma$ that scales the inverse of the baseline stiffness (so that $\sigma = 0$ corresponds to the rigid wing and $\sigma = 1$ to the baseline flexible configuration). The present strip theory results are compared with UVLM data of \citet{Murua2012}. The large wing tip displacements confirm the need for geometrically nonlinear analysis. The absence of a tail surface makes this configuration particularly sensitive to the coupling between structural deformation and flight dynamics, as the pitch stability must be maintained through the wing's own aerodynamic properties and the trailing-edge flap control.

The coupled framework has been applied to this configuration for several purposes in the authors' previous work, including nonlinear model order reduction achieving significant computational speedup compared to the full-order model while retaining accuracy in the gust response~\citep{Tantaroudas2015scitech}; worst-case gust identification using optimisation over the gust parameters to find the most severe structural response~\citep{DaRonch2013gust}; active gust load alleviation using both $\mathcal{H}_\infty$ robust control and model reference adaptive control (MRAC) strategies~\citep{Tantaroudas2014aviation, Tantaroudas2017bookchapter}; and flutter suppression using nonlinear control laws validated in wind tunnel tests~\citep{DaRonch2014flutter, Papatheou2013ifasd, Fichera2014isma}.

\subsection{Gust Response Simulation}
\label{sec:gust_response}

The gust response capability of the framework is demonstrated by subjecting the HALE aircraft to a 1-minus-cosine discrete gust at the trim condition ($U_\infty = 25$ m/s, $\rho_\infty = 0.0889$ kg/m$^3$)~\citep{Tantaroudas2015scitech}.

The gust penetration effect (Section~\ref{sec:gust_penetration}) is included, so the outboard strips encounter the gust later than the inboard strips due to the forward sweep of the gust front relative to the wing leading edge. The time history of the wing root bending moment and the wing tip displacement show oscillatory response with the dominant frequency corresponding to the first flapwise bending mode. The peak loads are well captured by the strip theory model when compared with UVLM results~\citep{DaRonch2013gust}.

The gust also excites the rigid-body phugoid and short-period modes, demonstrating the coupling between structural and flight dynamic responses. The phugoid oscillation is visible in the altitude and speed histories, while the short-period mode couples with the first structural bending mode at high dihedral angles~\citep{Patil2006, Hesse2014}.

\section{Discussion}
\label{sec:discussion}

The framework presented in this paper makes several modelling assumptions that define its range of applicability:

\paragraph{Two-dimensional strip aerodynamics.} The unsteady aerodynamic model assumes that each spanwise strip behaves as an independent two-dimensional aerofoil. This neglects three-dimensional effects (tip vortices, downwash variation along the span, aerodynamic interference between lifting surfaces) that are captured by panel methods or UVLM~\citep{Murua2012}. The strip theory provides good accuracy for high-aspect-ratio wings at low to moderate angles of attack (below 5--8 degrees), which is the typical operating range of HALE aircraft. For lower aspect ratios or higher incidence, a higher-fidelity aerodynamic model should be used. The modular nature of the framework allows replacement of the strip theory with UVLM or CFD without modifying the structural or flight dynamic components~\citep{DaRonch2012rom, Badcock2011}.

\paragraph{Thin-aerofoil theory.} The aerodynamic coefficients are based on flat-plate thin-aerofoil theory ($C_{L_\alpha} = 2\pi$), which assumes inviscid, incompressible, attached flow. Viscous and compressibility corrections can be introduced through modified $C_{L_\alpha}$ and profile drag coefficients. Flow separation and stall are not modelled, limiting the framework to pre-stall conditions.

\paragraph{Geometrically-exact beam.} The structural model captures arbitrarily large displacements and rotations but assumes that the cross-sectional dimensions are small compared to the beam length and that the cross-sectional shape does not change (no warping, no distortion). For thin-walled wing sections with significant warping, a refined cross-sectional analysis~\citep{Palacios2010intrinsic} should be used to compute the stiffness matrix $\mathbf{S}$.

\paragraph{Quasi-steady gravity.} The gravitational force is computed at the instantaneous deformed configuration and does not include time-lagged effects. This is appropriate for the slow deformations of HALE aircraft.

\paragraph{Computational efficiency.} The full-order HALE model with approximately 1,400 states can be integrated directly for time-domain simulation. For control design and optimisation applications that require repeated simulations, the model order reduction techniques developed in~\citep{DaRonch2013control, Tantaroudas2015scitech} can reduce the system to a small number of states while retaining the key dynamic behaviour, achieving speedups of up to 600$\times$.

\section{Conclusions}
\label{sec:conclusions}

A complete, self-contained mathematical framework for coupled aeroelastic-flight dynamic analysis of free-flying flexible aircraft has been presented. The framework integrates three tightly coupled disciplines. First, geometrically-exact nonlinear beam theory is used for structural dynamics, with explicit derivations of the kinematics (rotation tensor via the Rodrigues formula, force and moment strain measures $\boldsymbol{\gamma}$ and $\boldsymbol{\kappa}$), the constitutive law, the equations of motion (including mass, gyroscopic, and stiffness terms), and the finite element discretisation using two-noded displacement-based elements with six degrees of freedom per node. Second, unsteady two-dimensional strip aerodynamics based on Theodorsen thin-aerofoil theory is employed, with time-domain implementation through the Wagner and K\"ussner indicial functions using augmented aerodynamic states; the aerodynamic model includes section motion contributions, trailing-edge flap loads, and atmospheric gust loads, all formulated through Duhamel convolution integrals reduced to first-order ODEs. Third, quaternion-based rigid-body flight dynamics provides the Newton--Euler equations in the body frame and quaternion kinematic equations for singularity-free attitude propagation.

The coupled system is assembled into a first-order state-space form with a clearly defined state vector and Jacobian block structure. All coupling terms, including the effective velocity at each aerodynamic strip (linking rigid-body motion, structural deformation, and gust input), the coordinate transformations between frames, and the gust and control input matrices, have been derived explicitly.

The framework has been verified against established benchmarks: structural natural frequencies of the HALE wing are in close agreement with exact beam theory solutions and published results~\citep{Patil2001}; the flutter speed prediction (31.2 m/s) is within 6\% of UVLM-based results, with the difference attributable to the two-dimensional aerodynamic modelling; static aeroelastic trim deflections are in good agreement with CFD and UVLM results for the operating range of HALE aircraft; and the very flexible flying-wing configuration demonstrates the framework's ability to handle large deformations characteristic of HALE aircraft.

The state-space representation enables direct application of model order reduction for real-time simulation and optimisation~\citep{DaRonch2013control, Tantaroudas2015scitech}, linear and nonlinear control design for flutter suppression and gust load alleviation~\citep{Tantaroudas2014aviation, DaRonch2014flutter, Papatheou2013ifasd}, worst-case gust search for certification analysis~\citep{DaRonch2013gust}, and modular replacement of the aerodynamic model with higher-fidelity methods~\citep{Murua2012, DaRonch2012rom, Badcock2011}.

This paper provides a self-contained reference for researchers and engineers implementing coupled aeroelastic-flight dynamic analysis tools for very flexible aircraft. The detailed mathematical derivations, the explicit Jacobian structure, and the verification results should facilitate independent implementation and extension of the framework.

\section*{Acknowledgements}

Supported by EPSRC grant EP/I014594/1. The authors acknowledge Dr.\ H.\ Hesse and Dr.\ Y.\ Wang for assistance with the nonlinear beam code. The authors are grateful to Prof.\ K.J.\ Badcock and Prof.\ A.\ Da Ronch for guidance on nonlinear model order reduction, and to Prof.\ Rafael Palacios for guidance on the beam models.

\bibliographystyle{unsrtnat}
\bibliography{references}

\end{document}